\documentclass[11pt,a4paper]{article}
\usepackage{jheppub}
\usepackage{tikz}
\usetikzlibrary{quotes,arrows.meta}
\usepackage{dynkin-diagrams}
\usepackage{bm}
\usepackage{braket}
\usepackage{mathtools}
\usepackage{mleftright}
\usepackage{float}
\usepackage{array}
\usepackage{subcaption}
\allowdisplaybreaks[4]
\numberwithin{equation}{section}
\numberwithin{figure}{section}
\numberwithin{table}{section}
\DeclareMathOperator{\diag}{diag} %diagonal
\DeclareMathOperator{\tr}{Tr} %trace
\newcommand{\Z}{\mathbb{Z}}  %integer
\newcommand{\N}{\mathcal{N}}  %SUSY
\newcommand{\W}{\mathcal{W}}  %W algebra
\newcommand{\SW}{\mathcal{SW}}  %SUSY W algebra
\newcommand{\g}{\mathfrak{g}}  %Lie algebra
\renewcommand{\sl}{\mathfrak{sl}}  %special linear Lie algebra
\newcommand{\so}{\mathfrak{so}}  %special orthogonal Lie algebra
\renewcommand{\sp}{\mathfrak{sp}}  %symplectic Lie algebra
\newcommand{\osp}{\mathfrak{osp}}  %orthostmplectic Lie algebra
\newcommand{\svir}{\mathfrak{svir}}  %super-Virasoro algebra

\usepackage{stackengine,scalerel}
\newcommand{\textaltcolon}{%
  \ensurestackMath{%
    \stackon[1.5ex]
      {\scalebox{3.0}{$\boldsymbol{\circ}$}}
      {\scalebox{3.0}{$\boldsymbol{\circ}$}}%
  }%
}
\newcommand{\altcolon}{%
  \savestack\Tmp{\raisebox{-0.8pt}{\textaltcolon}}%
  \dp\Tmpcontent=\dimexpr\dp\Tmpcontent-0.8pt\relax%
  \mathrel{\scalerel*{\Tmp}{\scalebox{1.2}{$:$}}}%
}
\newcommand{\normord}[1]{\altcolon \! #1 \! \altcolon}

\preprint{TIT/HEP-714}

\title{\boldmath Free-field realizations of $\N = 1$ supersymmetric $\W$-algebras 
associated with Lie superalgebras $B (n - 1 , n)$ and $B (n , n)$}
\author{Naozumi Tanabe}

\affiliation{Department of Physics, Institute of Science Tokyo, \\
Tokyo, 152-8551, Japan}

\emailAdd{naotanabe.hep@gmail.com}

\abstract{
  We extend the recursive construction of free-field realizations proposed by Ide, Ito, 
  and Miyazaki to the $\N = 1$ supersymmetric $\W$-algebras $\SW B (n - 1 , n)$ and $\SW B (n , n)$, 
  associated with the Lie superalgebras $B (n - 1 , n)$ and $B (n , n)$, respectively. 
  We obtain explicit free-field realizations of $\SW B (0 , 1)$, $\SW B (1 , 1)$, $\SW B (1 , 2)$, 
  and $\SW B (2 , 2)$. 
  In particular, the realizations of $\SW B (1 , 2)$ and $\SW B (2 , 2)$ are new. 
  Using these realizations, we compute central charges, 
  conformal weights, and eigenvalues of the generating currents on free-field highest-weight states, 
  and construct the local integrals of motion $I_1$, $I_3$, and $I_5$. 
}


\begin{document}
\mleftright
\maketitle

\section{Introduction}
$\W$-algebras are higher-spin extensions of the Virasoro algebra in two-dimensional conformal field theory (CFT) 
\cite{zamolodchikov1985,fateev1987}. 
They play important roles in conformal field theory, integrable systems, 
representation theory, and supersymmetric gauge theories. 
For a Lie algebra $\g$ with a principal $\sl (2)$ embedding, 
the associated $\W$-algebra $\W \g$ can be constructed 
by quantum Drinfel'd--Sokolov (DS) reduction \cite{feigin1990,feigin1992}, 
which quantizes the classical DS construction \cite{drinfeld1985}; 
see also the review \cite{bouwknegt1993}. 
Its generating currents admit free-field realizations using Miura transformations or screening charges. 
In the latter approach, the $\W$-algebra consists of fields annihilated by a set of screening charges. 
Free field realizations have found applications in the AGT correspondence \cite{alday2010,nishioka2011} 
and the ODE/IM correspondence \cite{ito2023a,ito2025a,ide2026}. 
\par
Recently, a recursive construction of free-field realizations was developed in \cite{ide2026a}. 
When the Dynkin diagram of a rank-$r$ Lie algebra $\g_r$ 
contains a subdiagram corresponding to a rank-$(r - 1)$ Lie algebra $\g_{r - 1}$, 
the realization of $\W \g_r$ is constructed from that of $\W \g_{r - 1}$ 
by adjoining one free boson and imposing a zero commutation relation with the screening charges. 
Repeating this procedure gives a sequence of realizations 
\begin{equation}
  \W \g_1 \longrightarrow \W \g_2 \longrightarrow \cdots \longrightarrow \W \g_{r - 1} \longrightarrow \W \g_r . 
\end{equation}
Since the inherited zero commutation relations are automatically satisfied, 
only the zero commutation relation associated with the additional simple root needs to be imposed. 
Thus, the higher-rank problem reduces to finding combinations of lower-rank generators 
and one additional free boson that satisfy this relation. 
\par
In this paper, we extend this recursive construction of free-field realizations to 
$\N = 1$ supersymmetric (SUSY) $\W$-algebras associated with Lie superalgebras $B (n - 1 , n)$ and $B (n , n)$. 
We denote SUSY $\W$-algebras by $\SW$-algebras. 
Supersymmetric extensions of $\W$-algebras were investigated in early works 
\cite{inami1988,odake1989,blumenhagen1992,blumenhagen1992a}. 
More generally, Lie superalgebras admitting a principal $\osp (1 | 2)$ embedding 
and a purely odd simple root system include the following families \cite{kac1977,frappat1989}: 
\begin{equation}
  A (n , n - 1), \quad 
  B (n - 1 , n), \quad 
  B (n , n), \quad 
  C (2), \quad 
  D (n , n), \quad 
  D (n + 1 , n), \quad 
  D (2 , 1 ; \alpha) . 
\end{equation}
The corresponding $\SW$-algebras admit free-field realizations 
in terms of fermionic free superfields in the Cartan directions, 
with screenings associated with the odd simple roots. 
Explicit low-rank realizations have been constructed using 
SUSY Toda theories and Hamiltonian reduction \cite{evans1991,komata1991,ferreira1992,mallwitz1995}. 
A general screening realization of principal $\SW$-algebras at generic level was established in \cite{song2024}; 
see also \cite{song2024a}. 
Although the screening-kernel description characterizes these algebras, 
obtaining explicit higher-rank generating currents remains a separate computational problem. 
Here we address this problem for the two $B$-type families 
by constructing the new currents from lower-rank generators and one additional fermionic free superfield. 
\par
The rank-one algebra $\SW B (0 , 1)$, associated with $B (0 , 1) = \osp (1 | 2)$, 
coincides with the $\N = 1$ super-Virasoro algebra $\svir$ \cite{ferreira1992}. 
The higher-rank $\SW B$-algebras are therefore higher-spin extensions of $\svir$. 
They also arise as truncations of the $\N = 1$ $\W_\infty [\mu]$ algebra \cite{candu2013a}. 
The $B$-type families thus provide a natural setting for the recursive construction, 
with $\svir$ serving as the starting point. 
A related coset construction was studied in \cite{candu2013a}, where the chiral algebra of 
\begin{equation}
  \frac{\osp (1 | 2 n)_k \oplus \sp (2 n)_{- 1}}{\sp (2 n)_{k - 1}} 
\end{equation}
was conjectured to be isomorphic to $\SW B (n , n)$, providing a connection to $\N = 1$ minimal-model holography. 
No analogous coset realization of $\SW B (n - 1 , n)$ was found in that analysis. 
These results describe the algebras through their structure, truncations, and coset realizations. 
\par
To the best of our knowledge, explicit free-field expressions for the generating currents 
of the $\SW B$-algebras were previously available only up to $\SW B (1 , 1)$ \cite{mallwitz1995,diaz2016}. 
Here we extend the explicit free-field realizations to 
$\SW B (1 , 2)$ and $\SW B (2 , 2)$ using the recursive construction. 
These explicit expressions allow us to compute highest-weight charges 
and construct local integrals of motion (IoMs).
We obtain realizations of the first four members of the sequence 
\begin{equation}
  \begin{aligned}
    \begin{tikzpicture}[
        x=3.2cm, y=1.2cm,
        every node/.style={inner sep=3pt},
        >={Stealth}
      ]
      \node[anchor=east] at (-0.65,0)
        {$\SW B (n - 1 , n):$};
      \node[anchor=east] at (-0.65,-1)
        {$\SW B (n , n):$};
      \node (B01) at (0,0) {$\SW B (0 , 1)$};
      \node (B11) at (0.5,-1) {$\SW B (1 , 1)$};
      \node (B12) at (1,0) {$\SW B (1 , 2)$};
      \node (B22) at (1.5,-1) {$\SW B (2 , 2)$};
      \node[text opacity=0] (B23) at (2,0) {$\SW B (2 , 3)$};
      \node (dots) at (2.25,-0.5) {$\cdots$};
      \draw[->] (B01) -- (B11);
      \draw[->] (B11) -- (B12);
      \draw[->] (B12) -- (B22);
      \draw[->] (B22) -- (B23);
    \end{tikzpicture}
  \end{aligned}
  . 
\end{equation}
\par
The relation between simple root systems, Toda interactions, 
and SUSY has been studied systematically using Dynkin diagrams \cite{evans1997}. 
Quantum integrable structures provide a further motivation 
for studying explicit free-field realizations of $\SW$-algebras. 
For $\N = 1$ SCFT, these structures have been studied through monodromy matrices, transfer matrices, 
and Baxter's $Q$-operators associated with the SUSY KdV hierarchy \cite{kulish2004,kulish2005}. 
Related structures in coset CFTs, including the $\N = 1$ superconformal case, 
have been investigated using quantum ILW-type hierarchies and Bethe ansatz equations \cite{alfimov2015}. 
\par
The ODE/IM correspondence for SUSY affine Toda systems associated 
with affine Lie superalgebras admitting purely odd simple root systems was investigated in \cite{ito2022a}. 
Local IoMs of $\N = 1$ SCFT were computed numerically and compared with ODE/IM predictions in \cite{babenko2017}; 
see also \cite{babenko2019a}. 
More recently, WKB periods of a $C (2)^{(2)}$-type linear problem were matched with local IoM eigenvalues 
on Neveu--Schwarz highest-weight states through sixth order in the WKB expansion \cite{tanabe2026}. 
Our explicit construction of local IoMs for the higher-rank 
$\SW B$-algebras provides a basis for investigating analogous correspondences beyond the $\svir$ case. 
\par
Our main results are as follows. 
We obtain explicit free-field realizations of the first four $\SW B$-algebras in the sequence above 
and compute their central charges, conformal weights, and $\W$-charges of free-field highest-weight states. 
We also construct the local IoMs $I_1$, $I_3$, and $I_5$. 
For the $B (n , n)$ cases considered here, 
we fix $I_3$ using the odd affine screening and determine $I_5$ from $[I_3 , I_5] = 0$. 
For the $B (n - 1 , n)$ cases, we determine the densities directly from mutual commutativity. 
For $\SW B (1 , 2)$, this yields two rational branches, each giving a commuting pair $(I_3 , I_5)$. 
\par
We also obtain new results for $I_5$ of $\SW B (1 , 1)$ 
and for $I_3$ and $I_5$ of $\SW B (1 , 2)$ and $\SW B (2 , 2)$, 
extending the previous construction of $I_3$ for the $B (1 , 1)^{(1)}$ Toda theory \cite{penati1992}. 
\par
The remainder of this paper is organized as follows. 
Section~\ref{sec:SUSY_VA} introduces our conventions for $\N = 1$ superspace, superfield OPEs, and free superfields. 
Section~\ref{sec:free_field_realization} presents the recursive 
construction of free-field realizations of the $\SW B$-algebras up to rank four. 
Section~\ref{sec:W-ch} gives the conformal weights and $\W$-charges of free-field highest-weight states. 
Section~\ref{sec:IoM} constructs the first few local IoMs. 
We conclude in section~\ref{sec:conclusion}. 
Explicit expressions for the higher-spin currents of $\SW B (2 , 2)$ are collected in appendix~\ref{sec:SWB22}. 
\section{\boldmath $\N = 1$ superspace and OPEs}
\label{sec:SUSY_VA}
We consider the chiral algebra of two-dimensional $\N = 1$ superconformal field theories (SCFTs) 
in the Neveu--Schwarz (NS) sector. 
Thus, fermionic component fields are antiperiodic on the cylinder and have half-integer modes, 
whereas bosonic component fields have integer modes.
\par
A point in holomorphic superspace is denoted by $Z = (z , \theta)$, 
where $z$ is a bosonic (even) coordinate and $\theta$ is a fermionic (odd) coordinate satisfying $\theta^2 = 0$. 
A superfield admits the component expansion 
$\widehat{A} (Z) = A^{(0)} (z) + \theta A^{(1)} (z)$, 
where $A^{(0)}$ and $A^{(1)}$ have opposite Grassmann parities. 
Throughout this paper, a hat distinguishes a superfield from its component fields. 
\par
For two superspace points $Z_i = (z_i , \theta_i)$, we introduce the coordinate differences 
\begin{equation}
  z_{12} = z_1 - z_2 , \quad \theta_{12} = \theta_1 - \theta_2 , \quad Z_{12} = z_{12} - \theta_1 \theta_2 . 
\end{equation}
The covariant superderivative is defined by 
\begin{equation}
  D = \frac{\partial}{\partial \theta} + \theta \, \frac{\partial}{\partial z} , \quad 
  D^2 = \partial = \frac{\partial}{\partial z} . 
\end{equation}
\par
For two superfields $\widehat{A}$ and $\widehat{B}$, 
the singular part of their operator product expansion (OPE) is written as 
\begin{align}
  \widehat{A} (Z_1) \, \widehat{B} (Z_2) \sim \sum_{j \geq 0} 
  \left[ \frac{\theta_{12}}{Z_{12}^{j + 1}} \left( \! \widehat{A}_{(j | 0)} \widehat{B} \right) (Z_2) 
  + \frac{1}{Z_{12}^{j + 1}} \left( \! \widehat{A}_{(j | 1)} \widehat{B} \right) (Z_2) \right] . 
  \label{eq:OPE}
\end{align}
Here $\left( \! \widehat{A}_{(j | J)} \widehat{B} \right) (Z)$, 
with $J = 0 , 1$, denotes a local superfield appearing as an OPE coefficient. 
For each $j$, the coefficient superfields $\left( \! \widehat{A}_{(j | 0)} \widehat{B} \right)$ 
and $\left( \! \widehat{A}_{(j | 1)} \widehat{B} \right)$ have opposite Grassmann parities. 
For background on vertex (super)algebras and their SUSY extensions, 
see \cite{kac2001a} and \cite{heluani2007}, respectively. 
\par
As an example, consider the $\N = 1$ superconformal current, represented by the odd spin-$\frac{3}{2}$ superfield 
\begin{equation}
  \widehat{G} (Z) = G (z) + 2 \theta \, T (z) , 
\end{equation}
where $G$ and $T$ are the supercurrent and energy-momentum tensor, respectively. 
Its OPE is 
\begin{equation}
  \widehat{G} (Z_1) \, \widehat{G} (Z_2) 
  \sim \frac{1}{Z_{12}^3} \frac{2 c}{3} 
  + \frac{\theta_{12}}{Z_{12}^2} \, 3 \widehat{G} (Z_2) 
  + \frac{1}{Z_{12}} \, D_2 \widehat{G} (Z_2) 
  + \frac{\theta_{12}}{Z_{12}} \, 2 \partial_2 \widehat{G} (Z_2) , 
  \label{eq:svir-OPE}
\end{equation}
where $c$ is the central charge. 
This OPE is equivalent to the $\N = 1$ super-Virasoro algebra $\svir$ in components. 
\par
Following the $\N = 1$ SUSY vertex-algebra formalism of \cite{heluani2007}, 
it is convenient to encode the singular terms in \eqref{eq:OPE} into the SUSY $\varLambda$-bracket 
\begin{equation}
  \left[ \widehat{A}_\varLambda \widehat{B} \right] 
  = \sum_{j \geq 0} \frac{\lambda^j}{j !} 
  \left( \! \left( \! \widehat{A}_{(j | 0)} \widehat{B} \right) 
  + \chi \left( \! \widehat{A}_{(j | 1)} \widehat{B} \right) \! \right) , 
  \label{eq:lambda-bracket}
\end{equation}
where $\varLambda = (\lambda , \chi)$. 
$\lambda$ and $\chi$ are even and odd formal variables, respectively, satisfying $\chi^2 = - \lambda$. 
For example, the OPE \eqref{eq:svir-OPE} is encoded as 
\begin{equation}
  \left[ \widehat{G}_\varLambda \widehat{G} \right] 
  = \left(2 \partial + 3 \lambda + \chi \, D \right) \widehat{G} + \frac{c}{3} \, \lambda^2 \chi . 
  \label{eq:svir-algebra}
\end{equation}
The $\varLambda$-bracket computations in this paper were 
performed using the Mathematica package \texttt{Lambda} described in \cite{ekstrand2011}\footnote{
  The package can be downloaded from \url{https://github.com/joellernej/lambda}. 
}. 
\par
We denote by $\normord{\widehat{A} \widehat{B}}$ the normally ordered product, 
defined by the coincident-point limit of the regular part of the OPE. 
For nested products, the order of evaluation is indicated explicitly. 
\par
For a superfield $\widehat{A}$ of spin $\varDelta$, 
the $\varLambda$-bracket with the superconformal current takes the form 
\begin{equation}
  \left[ \widehat{G}_\varLambda \widehat{A} \right] 
  = \left( 2 \partial + 2 \varDelta \lambda + \chi \, D \right) \widehat{A} 
  + \mathcal{O} \left( \varLambda^2 \right) , 
  \label{eq:non-primary}
\end{equation}
where $\mathcal{O} \left( \varLambda^2 \right)$ denotes possible higher-order terms in the formal variables. 
The superfield $\widehat{A}$ is primary if these higher-order terms vanish, namely, 
\begin{equation}
  \left[ \widehat{G}_\varLambda \widehat{A} \right] 
  = \left( 2 \partial + 2 \varDelta \lambda + \chi \, D \right) \widehat{A} . 
  \label{eq:primary}
\end{equation}
\par
For the free-field realizations of $\SW$-algebras, we introduce the fermionic spin-$\frac{1}{2}$ superfield 
\begin{equation}
  \widehat{\varPsi} (Z) = \psi (z) + \theta \, \partial \phi (z) , 
\end{equation}
where $\phi$ and $\psi$ are a free boson and a free fermion, respectively. 
Their OPEs are 
\begin{equation}
  \partial \phi (z_1) \, \partial \phi (z_2) \sim \frac{\kappa}{z_{12}^2} , \quad 
  \psi (z_1) \, \psi (z_2) \sim \frac{\kappa}{z_{12}} , 
\end{equation}
where $\kappa$ is a constant. 
Equivalently, 
\begin{equation}
  \left[ \widehat{\varPsi}_\varLambda \widehat{\varPsi} \right] = \kappa \chi . 
  \label{eq:free-superfield}
\end{equation}
\section{Free-field realizations of \texorpdfstring{$\bm{\SW B (n - 1 , n)}$}{$\SW B (n - 1 , n)$} 
and \texorpdfstring{$\bm{\SW B (n , n)}$}{$\SW B (n , n)$} algebras}
\label{sec:free_field_realization}
In this section, we construct explicit free-field realizations 
of the $\SW B (n - 1 , n)$ and $\SW B (n , n)$ algebras up to rank four. 
We first introduce the free superfields and screening operators, 
and then construct the generating currents recursively. 

\subsection{Free superfields and screening operators}
The $\SW B (n - 1 , n)$ and $\SW B (n , n)$ algebras contain generating currents whose spins are known from 
the SUSY Toda field theories \cite{komata1991} and summarized in table~\ref{tab:N1_current_spin}. 
We construct these currents recursively from the generators of a lower-rank $\SW B$-algebra 
and one additional fermionic free superfield. 
The root-space decomposition underlying this construction is described below. 
\begin{table}[t]
  \centering 
  \begin{tabular}{ r | l }
    \hline 
    Lie superalgebra $\g$ & Spins of the $\W$-currents \\
    \hline 
    \quad & \quad \\[- 4mm]
    $B (n - 1 , n) = \osp (2 n - 1 | 2 n) \; (n \geq 1)$ & 
    $\dfrac{3}{2} ,\, 2 ,\, \dfrac{7}{2} ,\, 4 ,\, \dfrac{11}{2} ,\, 6 , \dots ,\, 
    2 n - \dfrac{5}{2} ,\, 2 n - 2 ,\, 2 n - \dfrac{1}{2}$ \\[2.5mm]
    \hline 
    \quad & \quad \\[- 4mm]
    $B (n , n) = \osp (2 n + 1 | 2 n) \; (n \geq 1)$ & 
    $\dfrac{3}{2} ,\, 2 ,\, \dfrac{7}{2} ,\, 4 ,\, \dfrac{11}{2} ,\, 6 , \dots ,\, 
    2 n - \dfrac{1}{2} ,\, 2 n$ \\[2.5mm]
    \hline 
  \end{tabular}
  \caption{Spins of the generating currents of the $\SW B (n - 1 , n)$ and $\SW B (n , n)$ algebras. 
  Each sequence terminates at the indicated largest spin. }
  \label{tab:N1_current_spin}
\end{table}
\par
Among the Lie superalgebras $B (m , n)$, those admitting a purely odd (fermionic) simple root system are 
\begin{equation}
  B (n - 1 , n) = \osp (2 n - 1 | 2 n) , \quad 
  B (n , n) = \osp (2 n + 1 | 2 n) . 
\end{equation}
For $B (m , n) = \osp (2 m + 1 | 2 n)$ with $m = n - 1$ or $m = n$, 
we describe the roots in terms of the basis vectors $\{ \epsilon_i \}_{i = 1}^{m}$ and $\{ \delta_j \}_{j = 1}^{n}$, 
whose inner products are normalized as 
\begin{equation}
  (\epsilon_i | \epsilon_j) = \delta_{i j} , \quad 
  (\delta_i | \delta_j) = - \delta_{i j} , \quad 
  (\epsilon_i | \delta_j) = 0 . 
\end{equation}
For convenience, we order the basis vectors as 
\begin{equation}
  e_a \coloneqq \left\{
    \begin{aligned}
      &\epsilon_a , &&& 1 &\leq a \leq m , \\
      &\delta_{a - m} , &&& m + 1 &\leq a \leq r , 
    \end{aligned}
  \right. 
  \label{eq:ordered-free-field-basis}
\end{equation}
where $r = m + n$ is the rank of $B (m , n)$. 
For each basis vector $e_a$, we introduce an odd free superfield $\widehat{\varPsi}_a$ satisfying 
\begin{equation}
  \left[ \widehat{\varPsi}_{a \, \varLambda} 
  \widehat{\varPsi}_b \right] = \eta_{a b} \, \chi , \quad 
  \eta_{a b} \coloneqq (e_a | e_b) 
  = \diag (\, \underbrace{1 , \dots , 1}_m ,\; \underbrace{- 1 , \dots , - 1}_n \,)_{a b} . 
  \label{eq:indexed-free-field-bracket}
\end{equation}
In vector notation, 
we write $\widehat{\varPsi} = \sum \limits_{a ,\, b = 1}^{r} \eta^{a b} \, \widehat{\varPsi}_a e_b$, 
where $\eta^{a b}$ is the inverse of the metric $\eta_{a b}$. 
\begin{figure}[t]
  \centering 
  \begin{dynkinDiagram}[
    labels={
        {\delta_1 - \epsilon_1},
        {\epsilon_1 - \delta_2},
        {\delta_2 - \epsilon_2},
        {\epsilon_2 - \delta_3},
        {\delta_{n - 1} - \epsilon_{n - 1}},
        {\epsilon_{n - 1} - \delta_n},
        {\delta_n}
      },
      edge length={1.7cm},
      root radius={0.15cm},
      text style/.style={scale=0.72},
      arrow width=2mm
      ]B{tttt.tt*}
  \end{dynkinDiagram}
  \caption{$B (n - 1 , n)$, 
  with the simple roots ordered from left to right as $\alpha_1 , \dots , \alpha_{2 n - 1}$. }
  \label{fig:B(n-1,n)}
\end{figure}
\begin{figure}[t]
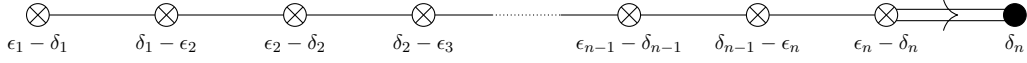

  \centering
  \begin{dynkinDiagram}[
      labels={
        {\epsilon_1 - \delta_1},
        {\delta_1 - \epsilon_2},
        {\epsilon_2 - \delta_2},
        {\delta_2 - \epsilon_3},
        {\epsilon_{n - 1} - \delta_{n - 1}},
        {\delta_{n - 1} - \epsilon_n},
        {\epsilon_n - \delta_n},
        {\delta_n}
      },
      edge length={1.7cm},
      root radius={0.15cm},
      text style/.style={scale=0.72},
      arrow width=2mm
      ]B{tttt.ttt*}
  \end{dynkinDiagram}
  \caption{$B (n , n)$, 
  with the simple roots ordered from left to right as $\alpha_1 , \dots , \alpha_{2 n}$. }
  \label{fig:B(n,n)}
\end{figure}
\par
We choose the purely odd simple root systems shown in figures~\ref{fig:B(n-1,n)} and \ref{fig:B(n,n)}. 
The simple roots $\alpha_i$ are ordered from left to right in each diagram. 
The crossed circles \dynkin[root radius={0.13cm}]A{t} represent isotropic odd simple roots, 
while the black node \dynkin[root radius={0.13cm}]A{*} represents the non-isotropic odd simple root $\delta_n$. 
For these root systems, we normalize the Weyl vector $\rho$ and the co-Weyl vector $\rho^\vee$ by\footnote{
  Our Weyl vector is related to the convention used, for example, 
  in \cite{candu2013a} by $\rho = 2 \rho_{\mathrm{std}}$, 
  where $(\rho_{\mathrm{std}} | \alpha_i) = \frac{1}{2} (\alpha_i | \alpha_i)$. 
  The normalization of $\rho^\vee$ is unchanged. 
}
\begin{equation}
  (\rho | \alpha_i) = (\alpha_i | \alpha_i) , \quad 
  \left( \rho^\vee \middle| \alpha_i \right) = 1 . 
  \label{eq:Weyl-vector-normalization}
\end{equation}
In this normalization, the Weyl and co-Weyl vectors are 
\begin{align}
  \rho (B (n - 1 , n)) &= - \sum_{i = 1}^{n - 1} \epsilon_i + \sum_{j = 1}^{n} \delta_j , &
  \rho^\vee (B (n - 1 , n)) 
  = \sum_{i = 1}^{n - 1} 2 (n - i) \epsilon_i &- \sum_{j = 1}^{n} (2 (n - j) + 1) \delta_j , \\
  \rho (B (n , n)) &= - \sum_{i = 1}^{n} \epsilon_i + \sum_{j = 1}^{n} \delta_j , &
  \rho^\vee (B (n , n)) 
  = \sum_{i = 1}^{n} 2 (n - i + 1) \epsilon_i &- \sum_{j = 1}^{n} (2 (n - j) + 1) \delta_j . 
\end{align}
\par
Introducing the coupling constant $\beta$, the superconformal current is 
\begin{equation}
  \widehat{G} (Z) = \; \normord{D \widehat{\varPsi} \cdot \widehat{\varPsi}} 
  + \; i \left( \! \beta \rho + \frac{1}{\beta} \, \rho^\vee \! \right) \cdot \partial \widehat{\varPsi} , 
  \label{eq:sconf_cur}
\end{equation}
which satisfies the $\svir$ relation \eqref{eq:svir-algebra} with central charge 
\begin{equation}
  c (B (m , n)) = \frac{3 r}{2} 
  + 3 \left( \! \beta \rho + \frac{1}{\beta} \rho^\vee 
  \middle| \beta \rho + \frac{1}{\beta} \, \rho^\vee \! \right) . 
  \label{eq:cent_ch}
\end{equation}
\par
For each simple root $\alpha_j$, we define the screening current 
\begin{equation}
  \widehat{S}_j (Z) \coloneqq \; 
  \normord{\exp \left( \! i \beta \int^Z \alpha_j \cdot \widehat{\varPsi} \left( Z' \right) \! \right)} . 
  \label{eq:screening-current}
\end{equation}
Here $\int^Z$ denotes integration with respect to the superderivative $D$, 
defined by $D \int^Z \widehat{\varPsi} (Z') = \widehat{\varPsi} (Z)$. 
With respect to the superconformal current \eqref{eq:sconf_cur}, 
the screening currents are primary superfields of spin $\frac{1}{2}$: 
\begin{equation}
  \left[ \widehat{G}_\varLambda \widehat{S}_j \right] 
  = (2 \partial + \lambda + \chi \, D) \widehat{S}_j . 
  \label{eq:screening-primary}
\end{equation}
The corresponding screening charge is 
\begin{equation}
  Q_j \coloneqq \oint dZ \, \widehat{S}_j (Z) , 
  \label{eq:screening-charge}
\end{equation}
where $dZ = dz \, d \theta$. 
The screening current $\widehat{S}_j$ is even, 
whereas the screening charge $Q_j$ is odd because of the Berezin integration. 
Commutators involving $Q_j$ are understood as supercommutators. 
The $\varLambda$-bracket of a free superfield with the screening current is 
\begin{equation}
  \left[ \widehat{\varPsi}_{a \, \varLambda} \widehat{S}_j \right] = i \beta (e_a | \alpha_j) \, \widehat{S}_j . 
  \label{eq:Psi-screening}
\end{equation}
The generating $\W$-currents satisfy the zero commutation relations 
\begin{equation}
  \left[ Q_j , \widehat{W}_s \right] 
  = \left[ \widehat{S}_{j \, \varLambda} \widehat{W}_s \right] \bigg|_{\lambda = \chi = 0} 
  = \left( \! \widehat{S}_{j \, (0 | 0)} \widehat{W}_s \! \right) = 0 , \quad j = 1 , \dots , r . 
  \label{eq:S_W}
\end{equation}
For generic $\beta$, the screening realization of the $\SW B (m , n)$ algebra is therefore given by 
\begin{equation}
  \SW B (m , n) = \bigcap_{j = 1}^{r} \ker Q_j . 
  \label{eq:SWB-kernel}
\end{equation}
Consequently, the contour integrals of the $\W$-currents 
supercommute with all screening charges \cite{feigin1992,genra2017,song2024,song2024a}. 

\subsection[Recursive construction of $\SW B (n - 1 , n)$ and $\SW B (n , n)$ algebras]
{\boldmath Recursive construction of $\SW B (n - 1 , n)$ and $\SW B (n , n)$ algebras}
\label{sec:SWB(m,n)}
We explicitly construct the first four algebras in the sequence 
\begin{equation}
  \begin{aligned}
    \begin{tikzpicture}[
        x=3.2cm, y=1.2cm,
        every node/.style={inner sep=3pt},
        >={Stealth}
      ]
      \node[anchor=east] at (-0.65,0)
        {$\SW B (n - 1 , n):$};
      \node[anchor=east] at (-0.65,-1)
        {$\SW B (n , n):$};
      \node (B01) at (0,0) {$\SW B (0 , 1)$};
      \node (B11) at (0.5,-1) {$\SW B (1 , 1)$};
      \node (B12) at (1,0) {$\SW B (1 , 2)$};
      \node (B22) at (1.5,-1) {$\SW B (2 , 2)$};
      \node[text opacity=0] (B23) at (2,0) {$\SW B (2 , 3)$};
      \node (dots) at (2.25,-0.5) {$\cdots$};
      \draw[->] (B01) -- (B11);
      \draw[->] (B11) -- (B12);
      \draw[->] (B12) -- (B22);
      \draw[->] (B22) -- (B23);
    \end{tikzpicture}
  \end{aligned}
  . 
  \label{eq:recursive-construction}
\end{equation}
At each step, the rightmost Dynkin subdiagram reproduces, up to relabeling, 
the purely odd simple root system of the preceding algebra. 
We construct the new currents from the generators of the preceding 
algebra and one additional fermionic free superfield. 
The inherited zero commutation relations are then automatically satisfied, 
leaving only the additional zero commutation relation to be imposed. 
This is the SUSY extension of the recursive construction proposed in \cite{ide2026a}. 
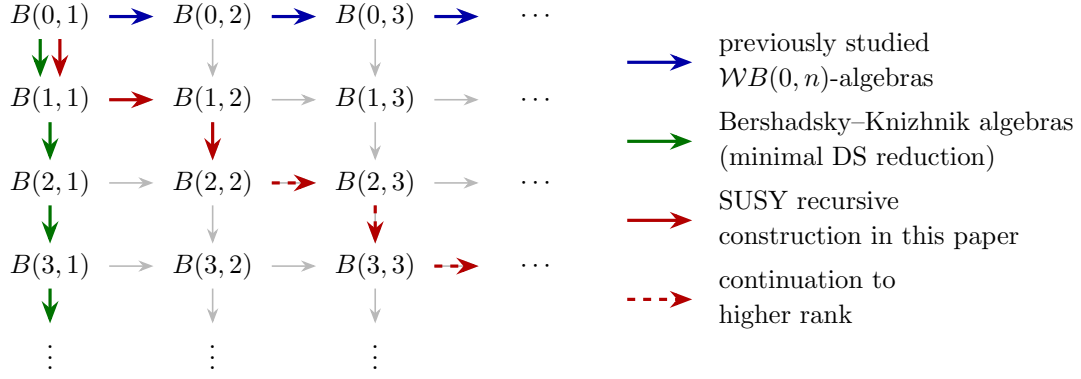
\begin{figure}[t]
  \centering
  \begin{tikzpicture}[
      x=2.15cm,
      y=-1.10cm,
      algebra/.style={
        minimum width=1.55cm,
        minimum height=0.58cm,
        inner sep=2pt
      },
      web/.style={
        ->,
        semithick,
        gray!55
      },
      known/.style={
        ->,
        very thick,
        blue!65!black
      },
      minimal/.style={
        ->,
        very thick,
        green!45!black
      },
      susy/.style={
        ->,
        very thick,
        red!70!black
      },
      continuation/.style={
        ->,
        very thick,
        dashed,
        red!70!black
      },
      every node/.style={font=\small},
      >={Stealth},
    ]
    % Algebra nodes
    \node[algebra] (B01) at (0,0) {$B(0,1)$};
    \node[algebra] (B02) at (1,0) {$B(0,2)$};
    \node[algebra] (B03) at (2,0) {$B(0,3)$};
    \node[algebra] (B11) at (0,1) {$B(1,1)$};
    \node[algebra] (B12) at (1,1) {$B(1,2)$};
    \node[algebra] (B13) at (2,1) {$B(1,3)$};
    \node[algebra] (B21) at (0,2) {$B(2,1)$};
    \node[algebra] (B22) at (1,2) {$B(2,2)$};
    \node[algebra] (B23) at (2,2) {$B(2,3)$};
    \node[algebra] (B31) at (0,3) {$B(3,1)$};
    \node[algebra] (B32) at (1,3) {$B(3,2)$};
    \node[algebra] (B33) at (2,3) {$B(3,3)$};
    % Horizontal and vertical continuation symbols
    \node[algebra] (R0) at (3,0) {$\cdots$};
    \node[algebra] (R1) at (3,1) {$\cdots$};
    \node[algebra] (R2) at (3,2) {$\cdots$};
    \node[algebra] (R3) at (3,3) {$\cdots$};
    \node[algebra] (D1) at (0,4) {$\vdots$};
    \node[algebra] (D2) at (1,4) {$\vdots$};
    \node[algebra] (D3) at (2,4) {$\vdots$};
    % Full B(m,n) web
    \draw[web] (B01) -- (B02);
    \draw[web] (B02) -- (B03);
    \draw[web] (B03) -- (R0);
    \draw[web] (B11) -- (B12);
    \draw[web] (B12) -- (B13);
    \draw[web] (B13) -- (R1);
    \draw[web] (B21) -- (B22);
    \draw[web] (B22) -- (B23);
    \draw[web] (B23) -- (R2);
    \draw[web] (B31) -- (B32);
    \draw[web] (B32) -- (B33);
    \draw[web] (B33) -- (R3);
    \draw[web] (B02) -- (B12);
    \draw[web] (B12) -- (B22);
    \draw[web] (B22) -- (B32);
    \draw[web] (B32) -- (D2);
    \draw[web] (B03) -- (B13);
    \draw[web] (B13) -- (B23);
    \draw[web] (B23) -- (B33);
    \draw[web] (B33) -- (D3);
    % Previously studied W B(0,n) series
    \draw[known] (B01) -- (B02);
    \draw[known] (B02) -- (B03);
    \draw[known] (B03) -- (R0);
    % B(m,1) series associated with highest-root sl(2) embeddings.
    \draw[minimal] ([xshift=-1.3mm]B01.south) -- ([xshift=-1.3mm]B11.north);
    \draw[minimal] (B11) -- (B21);
    \draw[minimal] (B21) -- (B31);
    \draw[minimal] (B31) -- (D1);
    % SUSY recursive construction obtained in this paper
    \draw[susy] ([xshift=1.3mm]B01.south) -- ([xshift=1.3mm]B11.north);
    \draw[susy] (B11) -- (B12);
    \draw[susy] (B12) -- (B22);
    % Expected continuation of the SUSY sequence
    \draw[continuation] (B22) -- (B23);
    \draw[continuation] (B23) -- (B33);
    \draw[continuation] (B33) -- (R3);
    % Legend to the right of the web
    \draw[known] (3.55,0.55) -- (3.95,0.55);
    \node[anchor=west,text width=4.2cm,align=left] at (4.05,0.55)
      {previously studied\\$\W B (0 , n)$-algebras};
    \draw[minimal] (3.55,1.5) -- (3.95,1.5);
    \node[anchor=west,text width=5.0cm,align=left] at (4.05,1.5)
      {Bershadsky--Knizhnik algebras\\
      (minimal DS reduction)};
    \draw[susy] (3.55,2.45) -- (3.95,2.45);
    \node[anchor=west,text width=4.2cm,align=left] at (4.05,2.45)
      {SUSY recursive\\construction in this paper};
    \draw[continuation] (3.55,3.4) -- (3.95,3.4);
    \node[anchor=west,text width=4.2cm,align=left] at (4.05,3.4)
      {continuation to\\higher rank};
  \end{tikzpicture}
  \caption{The web of $\W$-algebras associated with the Lie superalgebras $B (m , n)$. }
  \label{fig:B-web}
\end{figure}
\par
The Lie superalgebras $B (m , n)$ can be arranged in the two-dimensional web shown in figure~\ref{fig:B-web}. 
A horizontal step increases $n$ by one and corresponds to adjoining a $\delta$-direction, 
whereas a vertical step increases $m$ by one and corresponds to adjoining an $\epsilon$-direction. 
The upper boundary $B (0 , n) = \osp (1 | 2 n)$ corresponds 
to the previously studied $\W$-superalgebras $\W B (0 , n)$ \cite{lukyanov1990,ito1992a,genra2017}. 
Here a $\W$-superalgebra means a $\Z_2$-graded $\W$-algebra 
and should be distinguished from an $\N = 1$ SUSY $\W$-algebra, which we denote by $\SW$. 
At rank one, $\W B (0 , 1)$ coincides with the $\N = 1$ super-Virasoro algebra $\svir$. 
The Lie superalgebras $B (m , 1) = \osp (2 m + 1 | 2)$ along the left boundary
give rise to the $\W$-superalgebras known as the Bershadsky--Knizhnik algebras 
\cite{bershadsky1986,knizhnik1986}, which we denote here by $\W B (m , 1)$. 
These arise from minimal DS reduction, 
with the $\sl (2)$ embedding associated with the highest root of the $\sp (2)$ factor. 
The highlighted staircase consists of the two families $B (n - 1 , n)$ and $B (n , n)$, 
which admit purely odd simple root systems and give rise to the $\N = 1$ $\SW$-algebras studied in this paper. 

\subsubsection[$\SW B (0 , 1) = \svir$ algebra]{\boldmath $\SW B (0 , 1) = \svir$ algebra}
\label{sec:SWB(0,1)}
We begin with the rank-one algebra $\SW B (0 , 1)$, 
which coincides with the $\N = 1$ $\svir$. 
The single simple root of $B (0 , 1) = \osp (1 | 2)$ is non-isotropic and odd: 
\begin{equation}
  \alpha_1 = e_1 = \delta_1 , 
  \qquad 
  \begin{dynkinDiagram}[
      labels*={{\delta_1}},
      root radius={0.15cm},
      text style/.style={scale=0.85}
    ]B{*}
  \end{dynkinDiagram}
  . 
  \label{eq:dynkin-B01}
\end{equation}
For $B (0 , 1)$, the Weyl and co-Weyl vectors reduce to 
\begin{equation}
  \rho (B (0 , 1)) = e_1 , \quad \rho^\vee (B (0 , 1)) = - e_1 . 
  \label{eq:Weyl_B01}
\end{equation}
Substituting these expressions into \eqref{eq:sconf_cur}, we obtain 
\begin{equation}
  \widehat{G} = - \normord{D \widehat{\varPsi}_1 \, \widehat{\varPsi}_1} 
  + \; i \left( \! \beta - \frac{1}{\beta} \! \right) \partial \widehat{\varPsi}_1 . 
\end{equation}
The corresponding central charge is 
\begin{equation}
  c (B (0 , 1)) = \frac{15}{2} - 3 \left( \! \beta^2 + \frac{1}{\beta^2} \! \right) . 
  \label{eq:cc_B01}
\end{equation}
This free-field realization of $\svir$ serves as the starting point for the recursive construction below. 

\subsubsection[$\SW B (1 , 1)$ algebra]{\texorpdfstring{$\bm{\SW B (1 , 1)}$}{$\SW B (1 , 1)$} algebra}
\label{sec:SWB(1,1)}
The $\SW B (1 , 1)$ algebra is generated by the superconformal current $\widehat{G}$ 
and the $\W$-current $\widehat{W}_2$. 
This algebra has been studied in 
\cite{komata1991,ferreira1992,penati1992,blumenhagen1992a,mallwitz1995,gepner2001,diaz2016,fiset2020,robbins2024a}. 
We construct the $\SW B (1 , 1)$ algebra from the $\SW B (0 , 1)$ algebra. 
The simple roots of $B (1 , 1)$ are expressed
in terms of the root of $B (0 , 1)$ and an additional orthogonal direction as 
\begin{equation}
  \alpha_1 (B (1 , 1)) = e_1 - e_2 , \qquad 
  \alpha_2 (B (1 , 1)) = \alpha_1 (B (0 , 1)) , 
\end{equation}
where $e_1 = \epsilon_1$ is a unit vector orthogonal to the root space of $B (0 , 1)$. 
The purely odd simple root system is represented by 
\begin{equation}
  \begin{dynkinDiagram}[
      labels*={
        {\epsilon_1 - \delta_1},
        {\delta_1}
      },
      edge length={1.5cm},
      root radius={0.15cm},
      text style/.style={scale=0.85},
      arrow width=2mm
    ]B{t*}
    \dynkinBrace[B (0 , 1)]22
  \end{dynkinDiagram}
  , 
  \label{eq:dynkin-B11}
\end{equation}
where the brace indicates the $B (0 , 1)$ subdiagram. 
The Weyl and co-Weyl vectors of $B (1 , 1)$ are 
\begin{equation}
  \rho (B (1 , 1)) = - e_1 + \rho (B (0 , 1)) , \quad 
  \rho^\vee (B (1 , 1)) = 2 e_1 + \rho^\vee (B (0 , 1)) . 
  \label{eq:Weyl_B11}
\end{equation}
It follows from \eqref{eq:sconf_cur} that the superconformal current of $\SW B (1 , 1)$ is 
\begin{equation}
  \widehat{G} = \widehat{G}^{B (0 ,\, 1)} \; + \normord{D \widehat{\varPsi}_1 \, \widehat{\varPsi}_1} 
  - \; i \left( \! \beta - \frac{2}{\beta} \! \right) \partial \widehat{\varPsi}_1 . 
\end{equation}
The corresponding central charge is 
\begin{equation}
  c (B (1 , 1)) = - 3 + \frac{9}{\beta^2} . 
  \label{eq:cc_B11}
\end{equation}
\par
To illustrate the recursive construction, we construct the additional spin-2 current $\widehat{W}_2$ in detail. 
The current $\widehat{G}^{B (0 ,\, 1)}$ satisfies $\left[ Q_2 , \widehat{G}^{B (0 ,\, 1)} \right] = 0$, 
while $\widehat{\varPsi}_1$ has a regular OPE with $\widehat{S}_2$ because $(e_1 | \alpha_2) = 0$. 
Thus, composites constructed from these fields and their derivatives 
automatically satisfy the inherited zero commutation relation $\left[ Q_2 , \widehat{W}_2 \right] = 0$. 
The general ansatz of spin two and even Grassmann parity is 
\begin{equation}
  \widehat{W}_2 = b_1 \normord{\widehat{G}^{B (0 ,\, 1)} \, \widehat{\varPsi}_1} 
  + \; b_2 \normord{D^2 \widehat{\varPsi}_1 \, \widehat{\varPsi}_1} 
  + \; b_3 \normord{D \widehat{\varPsi}_1 \, D \widehat{\varPsi}_1} 
  + \; b_4 \, D \widehat{G}^{B (0 ,\, 1)} 
  + b_5 \, D^3 \widehat{\varPsi}_1 . 
\end{equation}
Imposing the remaining zero commutation relation $\left[ Q_1 , \widehat{W}_2 \right] = 0$ 
and collecting independent normally ordered products gives 
\begin{equation}
  b_1 = - 2 i \beta (b_3 - b_4) , \quad 
  b_2 = - 2 (\beta^2 - 1) b_3 + (2 \beta^2 - 1) b_4 , \quad 
  b_5 = - i \left( \beta - \frac{2}{\beta} \right) b_3 . 
\end{equation}
Two independent coefficients, $b_3$ and $b_4$, remain. 
The primary condition \eqref{eq:primary} requires $\left( \! \widehat{G}_{(3 | 0)} \widehat{W}_2 \! \right) 
= \left( \! \widehat{G}_{(1 | 1)} \widehat{W}_2 \! \right) = 0$. 
For generic $\beta$, these conditions give 
\begin{equation}
  2 \left( \beta^2 - 2 \right) b_3 - \left( 2 \beta^2 - 1 \right) b_4 = 0 . 
\end{equation}
Fixing the overall normalization by $b_2 = 1$ then yields 
\begin{align}
  b_1 &= \frac{3 i \beta}{2 \beta^2 - 1} , 
  & 
  b_2 &= 1 , 
  & 
  b_3 &= - \frac{1}{2} , 
  & 
  b_4 &= - \frac{\beta^2 - 2}{2 \beta^2 - 1} , 
  & 
  b_5 &= \frac{i}{2} \left( \! \beta - \frac{2}{\beta} \! \right) . 
\end{align}
Together with $\widehat{G}$, this current gives the free-field realization of $\SW B (1 , 1)$. 

\subsubsection[$\SW B (1 , 2)$ algebra]{\boldmath $\SW B (1 , 2)$ algebra}
\label{sec:SWB(1,2)}
The $\SW B (1 , 2)$ algebra is generated by the superconformal current $\widehat{G}$ 
and the $\W$-currents $\widehat{W}_2 ,\, \widehat{W}_{\frac{7}{2}}$. 
We construct the $\SW B (1 , 2)$ algebra from the $\SW B (1 , 1)$ algebra. 
The simple roots of $B (1 , 2)$ are expressed 
in terms of the simple roots of $B (1 , 1)$ and its orthogonal direction as 
\begin{equation}
  \alpha_1 (B (1 , 2)) = - e_1 + e_2 , \qquad 
  \alpha_{i + 1} (B (1 , 2)) = \alpha_i (B (1 , 1)) \quad (i = 1 , 2) , 
\end{equation}
where the $B (1 , 1)$ root system is embedded by relabeling its $\delta_1$ as $\delta_2$, 
and $e_2 = \delta_1$ is the additional direction orthogonal to its root space. 
The purely odd simple root system is represented by 
\begin{equation}
  \begin{dynkinDiagram}[
      labels*={
        {\delta_1 - \epsilon_1},
        {\epsilon_1 - \delta_2},
        {\delta_2}
      },
      edge length={1.5cm},
      root radius={0.15cm},
      text style/.style={scale=0.85},
      arrow width=2mm
    ]B{tt*}
    \dynkinBrace[B (1 , 1)]23
  \end{dynkinDiagram}
  , 
  \label{eq:dynkin-B12}
\end{equation}
where the brace indicates the $B (1 , 1)$ subdiagram. 
The Weyl and co-Weyl vectors of $B (1 , 2)$ are 
\begin{equation}
  \rho (B (1 , 2)) = \rho (B (1 , 1)) + e_2 , \quad 
  \rho^\vee (B (1 , 2)) = \rho^\vee (B (1 , 1)) - 3 e_2 . 
  \label{eq:Weyl_B12}
\end{equation}
It follows from \eqref{eq:sconf_cur} that the superconformal current of $\SW B (1 , 2)$ is 
\begin{equation}
  \widehat{G} = \widehat{G}^{B (1 ,\, 1)} \; - \normord{D \widehat{\varPsi}_2 \, \widehat{\varPsi}_2} 
  + \; i \left( \! \beta - \frac{3}{\beta} \! \right) \partial \widehat{\varPsi}_2 . 
\end{equation}
The corresponding central charge is 
\begin{equation}
  c (B (1 , 2)) = \frac{33}{2} - 3 \left( \! \beta^2 + \frac{6}{\beta^2} \! \right) . 
  \label{eq:cc_B12}
\end{equation}
\par
Following the procedure illustrated in section~\ref{sec:SWB(1,1)}, 
we construct $\widehat{W}_2$ and $\widehat{W}_{\frac{7}{2}}$ from the generators of $\SW B (1 , 1)$ 
and the additional free superfield $\widehat{\varPsi}_2$. 
The inherited zero commutation relations with $Q_2$ and $Q_3$ are automatically satisfied. 
The remaining zero commutation relation $\left[ Q_1 , \widehat{W}_s \right] = 0$ 
and the primary condition \eqref{eq:primary} fix the coefficients up to an overall normalization for each current. 
\par
For the spin-2 current, we take the ansatz 
\begin{equation}
  \widehat{W}_2 = b_1 \widehat{W}_2^{B (1 ,\, 1)} 
  + b_2 \, D \widehat{G}^{B (1 ,\, 1)} 
  + b_3 \normord{\widehat{G}^{B (1 ,\, 1)} \, \widehat{\varPsi}_2} 
  + \; b_4 \normord{\partial \widehat{\varPsi}_2 \, \widehat{\varPsi}_2} 
  + \; b_5 \normord{D \widehat{\varPsi}_2 \, D \widehat{\varPsi}_2} 
  + \; b_6 \, D^3 \widehat{\varPsi}_2 . 
\end{equation}
For generic $\beta$, we fix the overall normalization by setting $b_1 = 2 \beta^2 - 1$. 
With this choice, we obtain 
\begin{align}
  b_1 &= 2 \beta^2 - 1 , 
  & 
  b_2 &= \beta^2 - 3 , 
  & 
  b_3 &= - 3 i \beta , 
  & 
  b_4 &= - 2 , 
  & 
  b_5 &= 1 , &
  b_6 &= - i \left( \! \beta - \frac{3}{\beta} \! \right) . 
\end{align}
\par
For the spin-$\frac{7}{2}$ current, we take the ansatz
\begin{align}
  \widehat{W}_{\frac{7}{2}} 
  &= c_1 \normord{\widehat{W}_2^{B (1 ,\, 1)} \, \widehat{G}^{B (1 ,\, 1)}} 
  + \; c_2 \, D^3 \widehat{W}_2^{B (1 ,\, 1)} 
  + c_3 \, D^4 \widehat{G}^{B (1 ,\, 1)} 
  + c_4 \normord{D \widehat{G}^{B (1 ,\, 1)} \, \widehat{G}^{B (1 ,\, 1)}} \notag \\
  &\quad + c_5 \normord{\; \normord{\widehat{W}_2^{B (1 ,\, 1)} \, D \widehat{\varPsi}_2} \widehat{\varPsi}_2} 
  + \; c_6 \normord{D^2 \widehat{W}_2^{B (1 ,\, 1)} \, \widehat{\varPsi}_2} 
  + \; c_7 \normord{D \widehat{W}_2^{B (1 ,\, 1)} \, D \widehat{\varPsi}_2} 
  + \; c_8 \normord{\widehat{W}_2^{B (1 ,\, 1)} \, D^2 \widehat{\varPsi}_2} \notag \\
  &\quad + c_9 \normord{\; \normord{D \widehat{G}^{B (1 ,\, 1)} \, D \widehat{\varPsi}_2} \widehat{\varPsi}_2} 
  + \; c_{10} \normord{\; \normord{\widehat{G}^{B (1 ,\, 1)} \, D^2 \widehat{\varPsi}_2} \widehat{\varPsi}_2} 
  + \; c_{11} \normord{\; \normord{\widehat{G}^{B (1 ,\, 1)} \, D \widehat{\varPsi}_2} D \widehat{\varPsi}_2} 
  \notag \\
  &\quad + c_{12} \normord{D^3 \widehat{G}^{B (1 ,\, 1)} \, \widehat{\varPsi}_2} 
  + \; c_{13} \normord{D^2 \widehat{G}^{B (1 ,\, 1)} \, D \widehat{\varPsi}_2} 
  + \; c_{14} \normord{D \widehat{G}^{B (1 ,\, 1)} \, D^2 \widehat{\varPsi}_2} 
  + \; c_{15} \normord{\widehat{G}^{B (1 ,\, 1)} \, D^3 \widehat{\varPsi}_2} \notag \\
  &\quad + c_{16} \normord{\; \normord{\; \normord{D \widehat{\varPsi}_2 \, 
  D \widehat{\varPsi}_2} D \widehat{\varPsi}_2} \widehat{\varPsi}_2} 
  + \; c_{17} \normord{\; \normord{D^3 \widehat{\varPsi}_2 \, D \widehat{\varPsi}_2} \widehat{\varPsi}_2} 
  + \; c_{18} \normord{\; \normord{D^2 \widehat{\varPsi}_2 \, D \widehat{\varPsi}_2} D \widehat{\varPsi}_2} \notag \\
  &\quad + c_{19} \normord{D^5 \widehat{\varPsi}_2 \, \widehat{\varPsi}_2} 
  + \; c_{20} \normord{D^4 \widehat{\varPsi}_2 \, D \widehat{\varPsi}_2} 
  + \; c_{21} \normord{D^3 \widehat{\varPsi}_2 \, D^2 \widehat{\varPsi}_2} 
  + \; c_{22} \, D^6 \widehat{\varPsi}_2 . 
\end{align}
For generic $\beta$, we fix the independent overall normalization 
by setting $c_1 = \left( 2 \beta^2 - 1 \right) \left( 5 \beta^2 - 18 \right)$. 
With this choice, we obtain 
\begin{align}
  c_1 &= \left( 2 \beta^2 - 1 \right) \left( 5 \beta^2 - 18 \right) , \quad 
  c_2 = \frac{14 \beta^6 - 67 \beta^4 + 69 \beta^2 - 36}{2 \beta^2} , \notag \\
  c_3 &= - \frac{32 \beta^{10} - 422 \beta^8 + 1697 \beta^6 - 1881 \beta^4 + 432}
  {8 \beta^6 - 58 \beta^4 + 48 \beta^2} , \quad 
  c_4 = \frac{20 \beta^8 - 257 \beta^6 + 1085 \beta^4 - 1707 \beta^2 + 756}{4 \beta^4 - 29 \beta^2 + 24} , \notag \\
  c_5 &= - 18 \beta^2 + 42 , \quad 
  c_6 = c_7 = \frac{i \left( 10 \beta^4 - 59 \beta^2 + 60 \right)}{2 \beta} , \quad 
  c_8 = - \frac{2 i \left( \beta^4 - 11 \beta^2 - 3 \right)}{\beta} , \notag \\
  c_9 &= - \frac{3 \left( 4 \beta^6 - 7 \beta^4 - 86 \beta^2 + 24 \right)}{8 \beta^4 - 58 \beta^2 + 48} , \quad 
  c_{10} = - \frac{9 \left( 4 \beta^6 - 11 \beta^4 - 90 \beta^2 + 72 \right)}
  {8 \beta^4 - 58 \beta^2 + 48} , \notag \\
  c_{11} &= - \frac{9 \left( \beta^4 +\beta^2 - 12 \right)}{4 \beta^4 - 29 \beta^2 + 24} , \quad 
  c_{12} = - \frac{i \left( \beta^4 +\beta^2 - 12 \right)}{2 \beta} , \quad 
  c_{13} = - \frac{2 i \left( \beta^4 +\beta^2 - 12 \right)}{\beta} , \notag \\
  c_{14} &= \frac{i \left( 28 \beta^8 - 157 \beta^6 - 407 \beta^4 + 2334 \beta^2 - 1368 \right)}
  {8 \beta^5 - 58 \beta^3 + 48 \beta} , \quad 
  c_{15} = \frac{3 i \left( 4 \beta^8 - 22 \beta^6 - 59 \beta^4 + 327 \beta^2 - 180 \right)}
  {\beta  \left( 4 \beta^4 - 29 \beta^2 + 24 \right)} , \notag \\
  c_{16} &= \frac{9 \left( \beta^4 +\beta^2 - 12 \right)}{4 \beta^4 - 29 \beta^2 + 24} , \quad 
  c_{17} = c_{18} = - \frac{9 i \left( \beta^2 - 3 \right)^2 \, \left( \beta^2 + 4 \right)}
  {\beta \left( 4 \beta^4 - 29 \beta^2 + 24 \right)} , \notag \\
  c_{19} &= - \frac{6 \beta^8 - 51 \beta^6 - 93 \beta^4 + 720 \beta^2 - 432}
  {8 \beta^6 - 58 \beta^4 + 48 \beta^2} , \quad 
  c_{20} = - \frac{9 \left( 2 \beta^8 - 11 \beta^6 - 25 \beta^4 + 168 \beta^2 - 144 \right)}
  {8 \beta^6 - 58 \beta^4 + 48 \beta^2} , \notag \\
  c_{21} &= \frac{9 \left( \beta^8 - 6 \beta^6 - 16 \beta^4 + 87 \beta^2 - 36 \right)}
  {\beta^2 \left( 4 \beta^4 - 29 \beta^2 + 24 \right)} , \quad 
  c_{22} = \frac{i \left( \beta^2 - 3 \right)^2 \left( \beta^6 - 7 \beta^4 - 38 \beta^2 + 24 \right)}
  {\beta^3 \left( 4 \beta^4 - 29 \beta^2 + 24 \right)} . 
\end{align}
The resulting currents $\widehat{G}$, $\widehat{W}_2$, 
and $\widehat{W}_{\frac{7}{2}}$ provide the input for the next step of the recursive construction. 

\subsubsection[$\SW B (2 , 2)$ algebra]{\boldmath $\SW B (2 , 2)$ algebra}
\label{sec:SWB(2,2)}
The $\SW B (2 , 2)$ algebra is generated by the superconformal current $\widehat{G}$ 
and the $\W$-currents $\widehat{W}_2 ,\, \widehat{W}_{\frac{7}{2}} ,\, \widehat{W}_4$. 
We construct the $\SW B (2 , 2)$ algebra from the $\SW B (1 , 2)$ algebra. 
The simple roots of $B (2 , 2)$ are expressed 
in terms of the simple roots of $B (1 , 2)$ and a new orthogonal direction as 
\begin{equation}
  \alpha_1 (B (2 , 2)) = e_1 - e_3 , \qquad 
  \alpha_{i + 1} (B (2 , 2)) = \alpha_i (B (1 , 2)) \quad (i = 1 , 2 , 3) , 
\end{equation}
where the $B (1 , 2)$ root system is embedded by relabeling its $\epsilon_1$ as $\epsilon_2$, 
and $e_1 = \epsilon_1$ is the additional direction orthogonal to its root space. 
The purely odd simple root system is represented by 
\begin{equation}
  \begin{dynkinDiagram}[
      labels*={
        {\epsilon_{1} - \delta_{1}},
        {\delta_{1} - \epsilon_2},
        {\epsilon_2 - \delta_2},
        {\delta_2}
      },
      edge length={1.5cm},
      root radius={0.15cm},
      text style/.style={scale=0.85},
      arrow width=2mm
    ]B{ttt*}
    \dynkinBrace[B (1 , 2)]24
  \end{dynkinDiagram}
  , 
  \label{eq:dynkin-B22}
\end{equation}
where the brace indicates the $B (1 , 2)$ subdiagram. 
The Weyl and co-Weyl vectors of $B (2 , 2)$ are 
\begin{equation}
  \rho (B (2 , 2)) = \rho (B (1 , 2)) - e_1 , \quad 
  \rho^\vee (B (2 , 2)) = \rho^\vee (B (1 , 2)) + 4 e_1 . 
  \label{eq:Weyl_B22}
\end{equation}
It follows from \eqref{eq:sconf_cur} that the superconformal current of $\SW B (2 , 2)$ is 
\begin{equation}
  \widehat{G} = \widehat{G}^{B (1 ,\, 2)} + \normord{D \widehat{\varPsi}_1 \, \widehat{\varPsi}_1} 
  - \; i \left( \! \beta - \frac{4}{\beta} \! \right) \partial \widehat{\varPsi}_1 . 
\end{equation}
The corresponding central charge is 
\begin{equation}
  c (B (2 , 2)) = - 6 + \frac{30}{\beta^2} . 
  \label{eq:cc_B22}
\end{equation}
\par
Following the same procedure, we construct the additional generating currents $\widehat{W}_2$, 
$\widehat{W}_{\frac{7}{2}}$, and $\widehat{W}_4$ from the generators of $\SW B (1 , 2)$ 
and the additional free superfield $\widehat{\varPsi}_1$. 
The inherited zero commutation relations with $Q_2$, $Q_3$, and $Q_4$ are automatically satisfied. 
The remaining zero commutation relation $\left[ Q_1 , \widehat{W}_s \right] = 0$ 
and the primary condition \eqref{eq:primary} fix each current up to an overall normalization. 
For spin four, an additional ambiguity remains: 
a primary composite of lower-spin currents and their derivatives can be added. 
\par
For the spin-2 current, we take the ansatz 
\begin{equation}
  \widehat{W}_2 = \; b_1 \widehat{W}_2^{B (1 ,\, 2)} 
  + \; b_2 \, D \widehat{G}^{B (1 ,\, 2)} 
  + \; b_3 \normord{\widehat{G}^{B (1 ,\, 2)} \, \widehat{\varPsi}_1} 
  + \; b_4 \normord{D^2 \widehat{\varPsi}_1 \, \widehat{\varPsi}_1} 
  + \; b_5 \normord{D \widehat{\varPsi}_1 \, D \widehat{\varPsi}_1} 
  + \; b_6 \, D^3 \widehat{\varPsi}_1 . 
\end{equation}
For generic $\beta$, we fix the overall normalization by setting $b_3 = 1$. 
With this choice, we obtain 
\begin{align}
  b_1 &= - \frac{i}{3 \beta} , 
  & 
  b_2 &= \frac{i}{3} \left( \! \beta - \frac{4}{\beta} \! \right) , 
  & 
  b_3 &= 1 , \notag \\
  b_4 &= - i \left( \! \frac{2 \beta}{3} - \frac{1}{\beta} \! \right) , 
  &
  b_5 &= \frac{i}{2} \left( \! \frac{2 \beta}{3} - \frac{1}{\beta} \! \right) , 
  &
  b_6 &= \frac{1}{6} \left( \! \beta - \frac{4}{\beta} \! \right) \! 
  \left( \! \frac{2 \beta}{3} - \frac{1}{\beta} \! \right) .
\end{align}
The explicit expressions for $\widehat{W}_{\frac{7}{2}}$ and $\widehat{W}_4$, 
including their coefficient functions, are collected in appendix~\ref{sec:SWB22}. 
There we fix the composite-field ambiguity of $\widehat{W}_4$ by setting $d_{10} = 0$. 

\section{\texorpdfstring{$\bm{\W}$}{$\W$}-charges}
\label{sec:W-ch}
Having constructed the generating currents of the $\SW B (n - 1 , n)$ and $\SW B (n , n)$ algebras, 
we now determine their eigenvalues on free-field highest-weight states. 
Throughout this section, we write $\SW B (m , n)$ with $m = n - 1$ or $m = n$, and set $r = m + n$. 
The highest-weight data of an $\SW \g$-algebra representation are specified 
by the central charge and the eigenvalues of the relevant zero modes of the component currents. 
We refer to these eigenvalues as the $\W$-charges \cite{ide2026a}. 
\par
In the free-field realization, a highest-weight state is created by a vertex operator. 
Let $q = (q_1 , \dots , q_r)$ denote the representation labels 
and let $\mu = \mu(q)$ be the corresponding free-field momentum. 
We choose $q_i = (\mu | \alpha_i)$, where $\alpha_i$ are the ordered simple roots. 
We define 
\begin{equation}
  \widehat{V}_\mu (Z) \coloneqq \; 
  \normord{\exp \left( \! i \int^Z \mu \cdot \widehat{\varPsi} \left( Z' \right) \! \right)} . 
  \label{eq:vertex-operator}
\end{equation}
We denote the state created by $\widehat{V}_\mu$ by $\ket{\mu}$. 
The positive modes of the generating currents annihilate $\ket{\mu}$, 
whereas the relevant zero modes act diagonally. 
The corresponding highest-weight module is generated by acting on $\ket{\mu}$ with the negative modes. 
\par
The conformal weight $\varDelta (\mu)$ is determined by the leading term 
of the $\varLambda$-bracket with the superconformal current: 
\begin{equation}
  \left[ \widehat{G}_\varLambda \widehat{V}_\mu \right]_{\mathrm{leading}} 
  = 2 \lambda \, \varDelta (\mu) \, \widehat{V}_\mu . 
  \label{eq:conf-wt}
\end{equation}
Here $\varDelta(\mu)$ is the $L_0$ eigenvalue. 
The factor of 2 follows from the normalization $\widehat{G} = G + 2 \theta T$. 
For an additional generating current $\widehat{W}_s$, the $\W$-charge $w_s (\mu)$ can be read off from 
the highest-degree term in the $\varLambda$-bracket with the vertex operator: 
\begin{equation}
  \left[ \widehat{W}_{s \, \varLambda} \widehat V_\mu \right]_{\mathrm{leading}} = \left\{
    \begin{aligned}
      \displaystyle \frac{\lambda^{s - \frac{1}{2}}}{\left( s - 1 / 2 \right) !} \, 
      &w_s (\mu) \, \widehat{V}_\mu , & s &\in \Z + \frac{1}{2} , \\
      \displaystyle \frac{\lambda^{s - 1} \chi}{(s - 1) !} \, &w_s (\mu) \, \widehat{V}_\mu , & s &\in \Z . 
    \end{aligned}
  \right. 
  \label{eq:W-ch}
\end{equation}
The factorials follow from the convention \eqref{eq:lambda-bracket}. 
In the NS sector, $w_s$ is the zero-mode eigenvalue of the bosonic component: the lower component for integer $s$ 
and the upper component for half-integer $s$. 
The fermionic components have no zero modes. 
\par
Motivated by the shifted Weyl symmetry of the $\W$-charges, we introduce the shifted momentum 
\begin{equation}
  \widetilde{\mu} \coloneqq \mu - \frac{1}{2} \left( \! \beta \rho + \frac{1}{\beta} \, \rho^\vee \! \right) . 
  \label{eq:shifted-momentum}
\end{equation}
The factor and sign in this definition follow from our conventions for 
the vertex operators and the superconformal current. 
The conformal weight is then 
\begin{equation}
  \varDelta (\mu) 
  = - \frac{1}{2} \left( \! \mu \middle| \mu - \beta \rho - \frac{1}{\beta} \, \rho^\vee \! \right) 
  = - \frac{1}{2} \left( \widetilde{\mu} \middle| \widetilde{\mu} \right) 
  + \frac{c (B (m , n)) - \frac{3}{2} \, r}{24} , 
  \label{eq:conf-wt-general}
\end{equation}
where the second equality follows from the central charge \eqref{eq:cent_ch}. 
\par
The bosonic subalgebra of $B (m , n) = \osp (2 m + 1 | 2 n)$ is 
\begin{equation}
  B (m , n)_{\bar{0}} = B_m \oplus C_n = \so (2 m + 1) \oplus \sp (2 n) . 
\end{equation}
Using the orthogonal basis $\{ e_a \}_{a = 1}^{r}$ introduced in \eqref{eq:ordered-free-field-basis}, 
we define the shifted momentum components by 
\begin{equation}
  \widetilde{q}_a (q)
  \coloneqq \left( e_a \middle| \widetilde{\mu} (q) \right) ,
  \qquad a = 1 , \dots , r .
  \label{eq:shifted-momentum-components}
\end{equation}
Since the metric in this basis is $\eta_{a b}$, the shifted momentum is 
\begin{equation}
  \widetilde{\mu} (q)
  = \sum_{a = 1}^{m} \widetilde{q}_a (q) \, e_a
  - \sum_{a = m + 1}^{r} \widetilde{q}_a (q) \, e_a .
  \label{eq:shifted-momentum-decomposition}
\end{equation}
We denote the first and second terms in \eqref{eq:shifted-momentum-decomposition} by 
$\widetilde{\mu}^{\so (2 m + 1)}$ and $\widetilde{\mu}^{\sp (2 n)}$, respectively. 
Following the trace convention of \cite{ide2026a}, 
we define the Casimir invariants associated with the two bosonic factors by 
\begin{align}
  \mathcal{C}_{2 k}^{\so (2 m + 1)} (q) 
  &\coloneqq \tr_{\bm{2 m + 1}} 
  \left( \widetilde{\mu}^{\so (2 m + 1)} \cdot H^{\so (2 m + 1)} \right)^{\! 2 k} 
  = 2 \sum_{a = 1}^{m} \left( \widetilde{q}_a (q) \right)^{2 k} , & k &= 1 , \dots , m ,  
  \label{eq:B-Casimir} \\
  \mathcal{C}_{2 k}^{\sp (2 n)} (q) 
  &\coloneqq \tr_{\bm{2 n}} \left( \widetilde{\mu}^{\sp (2 n)} \cdot H^{\sp (2 n)} \right)^{\! 2 k} 
  = 2 \sum_{a = m + 1}^{r} \left( \widetilde{q}_a (q) \right)^{2 k} , & k &= 1 , \dots , n , 
  \label{eq:C-Casimir}
\end{align}
where $H^{\so (2 m + 1)}$ and $H^{\sp (2 n)}$ denote the Cartan generators
in the $(2 m + 1)$-dimensional vector representation of $\so (2 m + 1)$
and the $2 n$-dimensional fundamental representation of $\sp (2 n)$, respectively. 
The traces of odd powers vanish in both representations. 
The algebraically independent invariants therefore have degrees 
$2 , 4 , \dots , 2 m$ for $B_m$ and $2 , 4 , \dots , 2 n$ for $C_n$. 
In terms of the quadratic invariants, the general conformal-weight formula
\eqref{eq:conf-wt-general} becomes
\begin{equation}
  \varDelta (\mu)
  = - \frac{1}{4} \, \mathcal{C}_2^{\so (2 m + 1)} (q)
  + \frac{1}{4} \, \mathcal{C}_2^{\sp (2 n)} (q)
  + \frac{c (B (m , n)) - \frac{3}{2} \, r}{24} . 
  \label{eq:conf-wt-Casimir}
\end{equation}

\subsection[$\SW B (0 , 1)$ algebra]{\texorpdfstring{$\bm{\SW B (0 , 1)}$}{$\SW B (0 , 1)$} algebra}
\label{sec:B(0,1)_W-ch}
For $B (0 , 1) = \osp (1 | 2)$, the bosonic subalgebra is $\osp (1 | 2)_{\bar{0}} = \sp (2) \simeq \sl (2)$. 
The free-field momentum associated with the representation label $q_1$ is $\mu (q_1) = - q_1 e_1$. 
Using \eqref{eq:Weyl_B01}, the shifted momentum \eqref{eq:shifted-momentum} becomes 
\begin{equation}
  \widetilde{\mu} (q_1) = - \widetilde{q}_1 \, e_1 , \quad 
  \widetilde{q}_1 \coloneqq q_1 + \frac{1}{2} \left( \! \beta - \frac{1}{\beta} \! \right) . 
  \label{eq:B01-shifted-momentum}
\end{equation}
The quadratic Casimir invariant of $\sp (2)$ is therefore
\begin{equation}
  \mathcal{C}_2^{\sp (2)} (q_1) = 2 \widetilde{q}_1^{\, 2} . 
  \label{eq:B01-Casimir}
\end{equation}
Using \eqref{eq:conf-wt-Casimir} and the central charge \eqref{eq:cc_B01}, the conformal weight is 
\begin{equation}
  \varDelta (q_1) = \frac{1}{4} \, \mathcal{C}_2^{\sp (2)} (q_1) + \frac{c (B (0 , 1)) - \frac{3}{2}}{24} . 
  \label{eq:B01-conformal-weight}
\end{equation}
The conformal weight is invariant under $\widetilde{q}_1 \mapsto - \widetilde{q}_1$, 
or equivalently $q_1 \mapsto - q_1 - \left( \beta - \beta^{- 1} \right)$. 

\subsection[$\SW B (1 , 1)$ algebra]{\texorpdfstring{$\bm{\SW B (1 , 1)}$}{$\SW B (1 , 1)$} algebra}
\label{sec:B(1,1)_W-ch}
For $B (1 , 1) = \osp (3 | 2)$, the bosonic subalgebra is 
$\osp (3 | 2)_{\bar{0}} = \so (3) \oplus \sp (2) \simeq \sl (2)_1 \oplus \sl (2)_2$. 
The free-field momentum associated with the representation labels $q_1$ and $q_2$ is 
\begin{equation}
  \mu (q_1 , q_2) = (q_1 + q_2) e_1 - q_2 e_2 . 
\end{equation}
Using the Weyl and co-Weyl vectors of $B (1 , 1)$, the shifted momentum \eqref{eq:shifted-momentum} becomes 
\begin{equation}
  \widetilde{\mu} (q_1 , q_2) = \widetilde{q}_1 e_1 - \widetilde{q}_2 e_2 , 
\end{equation}
where 
\begin{equation}
  \widetilde{q}_1 \coloneqq q_1 + q_2 + \frac{\beta}{2} - \frac{1}{\beta} , \quad 
  \widetilde{q}_2 \coloneqq q_2 + \frac{\beta}{2} - \frac{1}{2 \beta} . 
  \label{eq:B11-shifted-momentum}
\end{equation}
The quadratic Casimir invariants of the two bosonic factors are 
\begin{equation}
  \mathcal{C}_2^{\so (3)} (q_1 , q_2) = 2 \widetilde{q}_1^{\, 2} , \quad 
  \mathcal{C}_2^{\sp (2)} (q_1 , q_2) = 2 \widetilde{q}_2^{\, 2} . 
  \label{eq:B11-Casimirs}
\end{equation}
Using \eqref{eq:conf-wt-Casimir} and \eqref{eq:W-ch}, we obtain the conformal weight and the $\W$-charge: 
\begin{align}
  \varDelta (q_1 , q_2) &= - \frac{1}{4} \, \mathcal{C}_2^{\so (3)} (q_1 , q_2) 
  + \frac{1}{4} \, \mathcal{C}_2^{\sp (2)} (q_1 , q_2) + \frac{c (B (1 , 1)) - 3}{24} , 
  \label{eq:B11-G-charge} \\
  w_2 (q_1 , q_2) &= \frac{1}{4} \, \mathcal{C}_2^{\so (3)} (q_1 , q_2) 
  - \frac{\beta^2 - 2}{2 \left( 2 \beta^2 - 1 \right)} \, \mathcal{C}_2^{\sp (2)} (q_1 , q_2) 
  + \frac{\beta^2 - 2}{8 \left( 2 \beta^2 - 1 \right)} . 
  \label{eq:B11-W2-charge}
\end{align}
Both expressions are invariant under the independent reflections 
$\left( \widetilde{q}_1 , \widetilde{q}_2 \right) \mapsto \left( - \widetilde{q}_1 , \widetilde{q}_2 \right)$ 
and $\left( \widetilde{q}_1 , \widetilde{q}_2 \right) \mapsto \left( \widetilde{q}_1 , - \widetilde{q}_2 \right)$. 

\subsection[$\SW B (1 , 2)$ algebra]{\texorpdfstring{$\bm{\SW B (1 , 2)}$}{$\SW B (1 , 2)$} algebra}
\label{sec:B(1,2)_W-ch}
For $B (1 , 2) = \osp (3 | 4)$, the bosonic subalgebra is 
$\osp (3 | 4)_{\bar{0}} = \so (3) \oplus \sp (4) \simeq \sl (2) \oplus \sp (4)$. 
The free-field momentum is 
\begin{equation}
  \mu (q_1 , q_2 , q_3) = (q_2 + q_3) e_1 - (q_1 + q_2 + q_3) e_2 - q_3 e_3 . 
\end{equation}
The shifted momentum is written as 
\begin{equation}
  \widetilde{\mu} (q_1 , q_2 , q_3)
  = \widetilde{q}_1 e_1 - \widetilde{q}_2 e_2 - \widetilde{q}_3 e_3 , 
\end{equation}
where 
\begin{equation}
  \widetilde{q}_1 = q_2 + q_3 + \frac{\beta}{2} - \frac{1}{\beta} , \quad 
  \widetilde{q}_2 = q_1 + q_2 + q_3 + \frac{\beta}{2} - \frac{3}{2 \beta} , \quad 
  \widetilde{q}_3 = q_3 + \frac{\beta}{2} - \frac{1}{2 \beta} . 
  \label{eq:B12-shifted-momentum}
\end{equation}
The relevant Casimir invariants are
\begin{equation}
  \mathcal{C}_2^{\so (3)} = 2 \widetilde{q}_1^{\,2} , \quad
  \mathcal{C}_2^{\sp (4)} = 2 \left( \widetilde{q}_2^{\,2} + \widetilde{q}_3^{\,2} \right) , \quad
  \mathcal{C}_4^{\sp (4)} = 2 \left( \widetilde{q}_2^{\,4} + \widetilde{q}_3^{\,4} \right) .
  \label{eq:B12-Casimirs}
\end{equation}
Their common dependence on $q = (q_1 , q_2 , q_3)$ is suppressed below.
Using \eqref{eq:conf-wt-Casimir} and \eqref{eq:W-ch}, we obtain the conformal weight and the $\W$-charges: 
\begin{align}
  \varDelta (q) 
  &= - \frac{1}{4} \, \mathcal{C}_2^{\so (3)}
  + \frac{1}{4} \, \mathcal{C}_2^{\sp (4)}
  + \frac{c (B (1 , 2)) - \frac{9}{2}}{24} , 
  \label{eq:B12-conformal-weight} \\[3mm]
  w_2 (q) 
  &= \frac{5}{4} \, \mathcal{C}_2^{\so (3)}
  - \frac{1}{2} \, \mathcal{C}_2^{\sp (4)}
  + \frac{4 - \beta^2}{8} , 
  \label{eq:B12-W2-charge} \\[3mm]
  w_{\frac{7}{2}} (q) 
  &= \frac{3 \left( 10 \beta^4 - 59 \beta^2 + 60 \right)}{8 \left( 2 \beta^2 - 1 \right)} \, \mathcal{C}_4^{\sp (4)} 
  - \frac{3 \left( 4 \beta^2 - 3 \right) \left( 5 \beta^4 - 52 \beta^2 + 120 \right)}
  {8 \left( 4 \beta^4 - 29 \beta^2 + 24 \right)} \left( \! \mathcal{C}_2^{\so (3)} \! \right)^{\! 2} \notag \\
  &\quad - \frac{3 \left( 40 \beta^8 - 550 \beta^6 + 2179 \beta^4 - 2856 \beta^2 + 1296 \right)}
  {16 \left( 2 \beta^2 - 1 \right) \left( 4 \beta^4 - 29 \beta^2 + 24 \right)} 
  \left( \! \mathcal{C}_2^{\sp (4)} \! \right)^{\! 2} \notag \\
  &\quad + \frac{3 \left( 20 \beta^6 - 229 \beta^4 + 630 \beta^2 - 288 \right)}
  {8 \left( 4 \beta^4 - 29 \beta^2 + 24 \right)} \, \mathcal{C}_2^{\so (3)} \mathcal{C}_2^{\sp (4)} \notag \\
  &\quad - \frac{3 \left( 34 \beta^6 - 390 \beta^4 + 1133 \beta^2 - 618 \right)}
  {8 \left( 4 \beta^4 - 29 \beta^2 + 24 \right)} \, \mathcal{C}_2^{\so (3)} \notag \\
  &\quad + \frac{3 \left( 16 \beta^8 - 302 \beta^6 + 1154 \beta^4 - 1045 \beta^2 + 444 \right)}
  {16 \left( 2 \beta^2 - 1 \right) \left( 4 \beta^4 - 29 \beta^2 + 24 \right)} \, \mathcal{C}_2^{\sp (4)} \notag \\
  &\quad + \frac{3 \left( 20 \beta^{12} - 244 \beta^{10} 
  + 1048 \beta^8 - 1772 \beta^6 + 558 \beta^4 + 321 \beta^2 - 180 \right)}
  {32 \beta^2 \left( 2 \beta^2 - 1 \right) \left( 4 \beta^4 - 29 \beta^2 + 24 \right)} . 
  \label{eq:B12-W72-charge}
\end{align}
Unlike $w_2$, the charge $w_{\frac{7}{2}}$ also involves the quartic Casimir invariant of $\sp (4)$. 

\subsection[$\SW B (2 , 2)$ algebra]{\texorpdfstring{$\bm{\SW B (2 , 2)}$}{$\SW B (2 , 2)$} algebra}
\label{sec:B(2,2)_W-ch}
For $B (2 , 2) = \osp (5 | 4)$, the bosonic subalgebra is $\osp (5 | 4)_{\bar{0}} = \so (5) \oplus \sp (4)$. 
The free-field momentum is 
\begin{align}
  \mu (q_1 , q_2 , q_3 , q_4) = (q_1 + q_2 + q_3 + q_4) e_1 + (q_3 + q_4) e_2 - (q_2 + q_3 + q_4) e_3 - q_4 e_4 . 
\end{align}
The shifted momentum is written as 
\begin{equation}
  \widetilde{\mu} (q_1 , q_2 , q_3 , q_4)
  = \widetilde{q}_1 e_1 + \widetilde{q}_2 e_2
  - \widetilde{q}_3 e_3 - \widetilde{q}_4 e_4 ,
  \label{eq:B22-shifted-momentum}
\end{equation}
where
\begin{align}
  \widetilde{q}_1 &= q_1 + q_2 + q_3 + q_4 + \frac{\beta}{2} - \frac{2}{\beta} , & 
  \widetilde{q}_2 &= q_3 + q_4 + \frac{\beta}{2} - \frac{1}{\beta} , \notag \\
  \widetilde{q}_3 &= q_2 + q_3 + q_4 + \frac{\beta}{2} - \frac{3}{2 \beta} , & 
  \widetilde{q}_4 &= q_4 + \frac{\beta}{2} - \frac{1}{2 \beta} . 
\end{align}
The relevant trace Casimir invariants are
\begin{equation}
  \mathcal{C}_{2 k}^{\so (5)}
  = 2 \left( \! \widetilde{q}_1^{\, 2 k} + \widetilde{q}_2^{\, 2 k} \! \right) , \quad 
  \mathcal{C}_{2 k}^{\sp (4)}
  = 2 \left( \! \widetilde{q}_3^{\, 2 k} + \widetilde{q}_4^{\, 2 k} \! \right) , \quad k = 1 , 2 .
  \label{eq:B22-Casimirs}
\end{equation}
Their common dependence on $q = (q_1 , q_2 , q_3 , q_4)$ is suppressed below.
Using \eqref{eq:conf-wt-Casimir} and \eqref{eq:W-ch}, we obtain the conformal weight and the $\W$-charges: 
\begin{align}
  \varDelta (q) &= - \frac{1}{4} \, \mathcal{C}_2^{\so (5)} 
  + \frac{1}{4} \, \mathcal{C}_2^{\sp (4)} + \frac{c (B (2 , 2)) - 6}{24} , 
  \label{eq:B22-conformal-weight} \\[3mm]
  w_2 (q) &= - \frac{i \left( 2 \beta^2 - 3 \right)}{12 \beta} \, \mathcal{C}_2^{\so (5)} 
  + \frac{i \left( \beta^2 - 3 \right)}{6 \beta} \, \mathcal{C}_2^{\sp (4)} 
  - \frac{i \left( \beta^2 - 4 \right)}{12 \beta} , 
  \label{eq:B22-W2-charge} \\[3mm]
  w_{\frac{7}{2}} (q) 
  &= - \frac{3 \left( 4 \beta^2 - 25 \right) \left( \beta^2 - 40 \right)}{8} \, \mathcal{C}_4^{\so (5)} 
  + \frac{3 \left( 4 \beta^2 - 25 \right) \left( \beta^2 - 40 \right)}{8} \, \mathcal{C}_4^{\sp (4)} \notag \\
  &\quad + \frac{3 \left( \beta^2 + 2 \right) \left( 18 \beta^4 - 161 \beta^2 + 410 \right)}{16} 
  \left( \! \mathcal{C}_2^{\so (5)} \! \right)^{\! 2} 
  + \frac{3 \left( 18 \beta^6 - 128 \beta^4 + 235 \beta^2 - 260 \right)}{16} 
  \left( \! \mathcal{C}_2^{\sp (4)} \! \right)^{\! 2} \notag \\
  &\quad - \frac{3 \left( 36 \beta^6 - 253 \beta^4 + 323 \beta^2 + 560 \right)}{16} \, 
  \mathcal{C}_2^{\so (5)} \mathcal{C}_2^{\sp (4)} 
  + \frac{3 \left( 49 \beta^6 - 326 \beta^4 + 379 \beta^2 + 375 \right)}{16} \, 
  \mathcal{C}_2^{\so (5)} \notag \\
  &\quad - \frac{3 \left( 49 \beta^6 - 320 \beta^4 + 118 \beta^2 + 1215 \right)}{16} \, 
  \mathcal{C}_2^{\sp (4)} 
  + \frac{45 \left( 4 \beta^8 - 30 \beta^6 + 28 \beta^4 + 80 \beta^2 - 25 \right)}{32 \beta^2} , 
  \label{eq:B22-W72-charge} \\
  w_4 (q) &= - \frac{15 i \beta \left( \beta^2 - 4 \right) \left( 4 \beta^2 - 3 \right) 
  \left( 17 \beta^2 + 60 \right) \left( 67 \beta^2 - 300 \right)}{112} \, \mathcal{C}_4^{\so (5)} \notag \\
  &\quad + \frac{3 i \beta \left( 17 \beta^2 + 60 \right) \left( 67 \beta^2 - 300 \right) 
  \left( 10 \beta^6 - 70 \beta^4 + 157 \beta^2 - 120 \right)}{28 \left( 2 \beta^2 - 1 \right)} \, 
  \mathcal{C}_4^{\sp (4)} \notag \\
  &\quad - \frac{15 i \beta \left( \beta^2 - 4 \right) \left( 4 \beta^2 - 3 \right) 
  \left( 519 \beta^6 - 2042 \beta^4 - 2379 \beta^2 + 14940 \right)}{112} 
  \left( \! \mathcal{C}_2^{\so (5)} \! \right)^{\! 2} \notag \\
  &\quad - \frac{3 i \beta}{112 \left( 2 \beta^2 - 1 \right)} 
  \big( 20760 \beta^{12} - 222830 \beta^{10} + 649813 \beta^8 \notag \\
  &\qquad + 583017 \beta^6 - 4691400 \beta^4 + 3497760 \beta^2 + 1425600 \big) 
  \left( \! \mathcal{C}_2^{\sp (4)} \! \right)^{\! 2} \notag \\
  &\quad + \frac{3 i \beta \left( 4 \beta^2 - 3 \right) 
  \left( 2595 \beta^8 - 22600 \beta^6 + 39218 \beta^4 + 128640 \beta^2 - 352800 \right)}{56} \, 
  \mathcal{C}_2^{\so (5)} \mathcal{C}_2^{\sp (4)} \notag \\
  &\quad - \frac{3 i \beta \left( 4 \beta^2 - 3 \right) 
  \left( 9905 \beta^8 - 83268 \beta^6 + 134852 \beta^4 + 463120 \beta^2 - 1180800 \right)}{224} \, 
  \mathcal{C}_2^{\so (5)} \notag \\
  &\quad + \frac{3 i \beta}{56 \left( 2 \beta^2 - 1 \right)} 
  \big( 19810 \beta^{12} - 228316 \beta^{10} + 767951 \beta^8 \notag \\
  &\qquad + 313042 \beta^6 - 5756526 \beta^4 + 7811100 \beta^2 - 2019600 \big) 
  \mathcal{C}_2^{\sp (4)} \notag \\
  &\quad - \frac{3 i \beta}{448 \beta^2 \left( 2 \beta^2 - 1 \right)} 
  \big( 48450 \beta^{14} - 604315 \beta^{12} + 2453160 \beta^{10} \notag \\
  &\qquad - 971000 \beta^8 - 15085580 \beta^6 + 27965872 \beta^4 - 12022320 \beta^2 + 1425600 \big) . 
  \label{eq:B22-W4-charge}
\end{align}
In all four examples, 
the charges are invariant under the Weyl groups of the bosonic subalgebras acting on the shifted momenta, 
providing a consistency check of the free-field realizations. 
\section{Local integrals of motion}
\label{sec:IoM}
We now construct the first few local integrals of motion (IoMs) 
of the $\SW B$-algebras obtained in the previous sections. 
The spins of the local IoMs associated with
$\SW B (n - 1 , n)$ and $\SW B (n , n)$ are summarized in table~\ref{tab:IoM-spins}. 
Local IoMs of half-integer spin are absent from these commuting hierarchies \cite{penati1992,delduc1998}. 
\begin{table}[t]
  \centering
  \begin{tabular}{l | c}
    \hline
    Algebra & Spins of the local IoMs $I_k$ \\
    \hline
    $\SW B (n - 1 , n)$ & $1 ,\, 3 ,\, 5 ,\, \dots$ \\
    $\SW B (n , n)$     & $1 ,\, 3 ,\, 5 ,\, \dots$ \\
    \hline
  \end{tabular}
  \caption{Spins of the local IoMs $I_k$ associated with $\SW B (n - 1 , n)$ and $\SW B (n , n)$. }
  \label{tab:IoM-spins}
\end{table}
In this paper, we explicitly construct the first three local IoMs, $I_1$, $I_3$, and $I_5$. 
We define 
\begin{equation}
  I_k \coloneqq \oint dZ \, \widehat{J}_{k + \frac{1}{2}} (Z) , 
  \label{eq:IoM-definition}
\end{equation}
where $\widehat{J}_{k + \frac{1}{2}}$ is a local odd superfield of spin $k + \frac{1}{2}$. 
We take $\widehat{J}_{\frac{3}{2}} = \widehat{G}$, which gives 
\begin{equation}
  I_1 = \oint dz \, d \theta \, \widehat{G} (Z) = 2 \oint dz \, T (z) . 
\end{equation}
Thus, $I_1$ generates translations and automatically commutes with the other local IoMs. 
\par
Since the superspace integral of a total superderivative vanishes, the density is defined modulo 
\begin{equation}
  \widehat{J}_{k + \frac{1}{2}} \sim \widehat{J}_{k + \frac{1}{2}} + D \widehat{X}_k . 
  \label{eq:total_derivative}
\end{equation}
Ordinary total derivatives are included in this equivalence because $\partial = D^2$. 
Two local IoMs commute if and only if the zeroth product of their densities is a total superderivative: 
\begin{equation}
  [I_n , I_m] = 0 \quad \Longleftrightarrow \quad 
  \left[ \widehat{J}_{n + \frac{1}{2} \, \varLambda} \widehat{J}_{m + \frac{1}{2}} \right] 
  \bigg|_{\lambda = \chi = 0} 
  = \left( \! \widehat{J}_{n + \frac{1}{2} \, (0 | 0)} \widehat{J}_{m + \frac{1}{2}} \right) 
  = D \widehat{X}_{n ,\, m} , 
  \label{eq:conserved_law}
\end{equation}
where $\widehat{X}_{n ,\, m}$ is a local composite superfield constructed from the superconformal current, 
the generating $\W$-currents, and their derivatives. 
\par
The classical origin of these hierarchies is the Drinfel'd--Sokolov construction, 
which associates KdV-type integrable hierarchies with affine Lie algebras \cite{drinfeld1985}. 
Its SUSY extension in terms of loop superalgebras and manifest superspace 
was developed in \cite{delduc1998}. 
Quantum higher-spin conserved currents in $\N = 1$ SUSY Toda theories 
admitting a fermionic untwisted affine extension were studied in \cite{gualzetti1993}. 
\par
For ordinary simple Lie algebras, the local IoMs of the quantum affine Toda theory can be characterized 
by requiring them to commute with the screening charges associated 
with all simple roots of the corresponding affine Lie algebra \cite{feigin1992a,feigin1996}. 
The screenings associated with the finite simple roots define the $\W$-algebra, 
while the additional affine screening selects a commutative family of quantum KdV Hamiltonians. 
\begin{figure}[t]
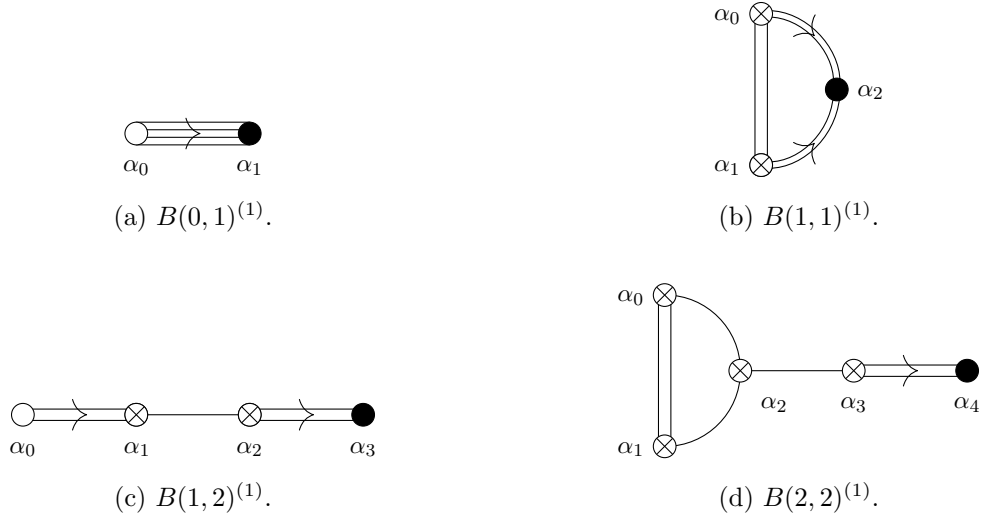

  \centering
  \begin{subfigure}[b]{0.47\textwidth}
    \centering
    \begin{dynkinDiagram}[
      labels={
        {\alpha_0},
        {\alpha_1}
      },
      affine mark={o},
      extended,
      reverse arrows,
      edge length={1.5cm},
      root radius={0.15cm},
      fold radius={1cm},
      text style/.style={scale=0.85},
      arrow width=2mm
    ]A[2]2
    \end{dynkinDiagram}
    \caption{$B (0 , 1)^{(1)}$.}
    \label{fig:B(0,1)(1)}
  \end{subfigure}
  \hfill
  \begin{subfigure}[b]{0.47\textwidth}
    \centering
    \begin{dynkinDiagram}[
      labels={
        {},
        {\alpha_2},
        {}
      },
      labels*={
        {\alpha_0},
        {},
        {\alpha_1}
      },
      affine mark={t},
      backwards,
      fold,
      double edges,
      edge/.append style={double distance=2pt},
      fold style/.append style={-},
      edge length={1.5cm},
      root radius={0.15cm},
      fold radius={1cm},
      reverse arrows,
      text style/.style={scale=0.85},
      arrow width=2mm
    ]D[2]3
    \dynkinRootMark{t}2
  \end{dynkinDiagram}
    \caption{$B (1 , 1)^{(1)}$.}
    \label{fig:B(1,1)(1)}
  \end{subfigure}
  \par\vspace{5mm}
  \begin{subfigure}[b]{0.47\textwidth}
    \centering
    \begin{dynkinDiagram}[
      labels={
        {\alpha_0},
        {\!\!\!\!\!\!\!\! \alpha_1},
        {\alpha_2},
        {\alpha_3}
      },
      affine mark={o},
      edge length={1.5cm},
      root radius={0.15cm},
      fold radius={1cm},
      reverse arrows,
      text style/.style={scale=0.85},
      arrow width=2mm
    ]A[2]6
    \dynkinRootMark{t}1
    \dynkinRootMark{t}2
  \end{dynkinDiagram}
    \caption{$B (1 , 2)^{(1)}$.}
    \label{fig:B(1,2)(1)}
  \end{subfigure}
  \hfill
  \begin{subfigure}[b]{0.47\textwidth}
    \centering
    \begin{dynkinDiagram}[
      labels={
        {\alpha_0},
        {\alpha_1},
        {\alpha_2},
        {\alpha_3},
        {\alpha_4}
      },
      affine mark={t},
      fold,
      double edges,
      fold style/.append style={-},
      edge length={1.5cm},
      root radius={0.15cm},
      fold radius={1cm},
      text style/.style={scale=0.85},
      arrow width=2mm
    ]B[1]{ttt*}
  \end{dynkinDiagram}
    \caption{$B (2 , 2)^{(1)}$.}
    \label{fig:B(2,2)(1)}
  \end{subfigure}
  \caption{Affine Dynkin diagrams relevant to the local integrable hierarchies. 
  The affine root $\alpha_0$ is even (\dynkin[root radius={0.13cm}]A{o}) 
  for $B (0 , 1)^{(1)}$ and $B (1 , 2)^{(1)}$, 
  whereas it is odd (\dynkin[root radius={0.13cm}]A{t}) for $B (1 , 1)^{(1)}$ and $B (2 , 2)^{(1)}$. 
  Only in the latter two cases do we impose commutativity with the affine screening charge to determine $I_3$. }
  \label{fig:B(m,n)(1)}
\end{figure}
\par
An analogous construction can be applied directly to the $\SW B (n , n)$ algebras considered here. 
For our choice of the purely odd simple root system, the affine root is 
\begin{equation}
  \alpha_0 \left( \! B (n , n)^{(1)} \right) = - \epsilon_1 - \delta_1 , 
  \label{eq:B(n,n)_aff_root}
\end{equation}
which is odd. 
We may therefore introduce the corresponding screening current 
$\widehat{S}_0$ in the same way as the finite-root screenings. 
Since every $\SW B (n , n)$ current already lies in the kernels of the finite screening charges, 
we impose the commutativity condition with the affine screening charge:
\begin{equation}
  [Q_0 , I_k] = 0 \quad \Longleftrightarrow \quad 
  \left[ \widehat{S}_{0 \, \varLambda} \widehat{J}_{k + \frac{1}{2}} \right] \bigg|_{\lambda = \chi = 0} 
  = \left( \! \widehat{S}_{0 \, (0 | 0)} \widehat{J}_{k + \frac{1}{2}} \right) 
  = D \! \normord{\widehat{S}_0 \widehat{P}_k} \, , 
  \label{eq:integrable_perturbation}
\end{equation}
where $\widehat{P}_k$ is a local differential polynomial in the free superfields. 
In practice, we fix $I_3$ using the commutativity condition \eqref{eq:integrable_perturbation}
and then determine $I_5$ by imposing mutual commutativity, $[I_3 , I_5] = 0$. 
\par
The situation is different for the $B (n - 1 , n)$ series. 
In this case, the affine root is 
\begin{equation}
  \alpha_0 \left( \! B (n - 1 , n)^{(1)} \right) = - 2 \delta_1 , 
  \label{eq:B(n-1,n)_aff_root}
\end{equation}
which is even and falls outside the purely odd screening framework used here. 
We therefore determine the local densities directly from mutual commutativity \eqref{eq:conserved_law}. 
\par
The first four affine Dynkin diagrams considered below are summarized in figure~\ref{fig:B(m,n)(1)}. 
The left column corresponds to the $B (n - 1 , n)$ series, 
whereas the right column corresponds to the $B (n , n)$ series. 

\subsection[$\SW B (0 , 1)$ algebra]{\texorpdfstring{$\bm{\SW B (0 , 1)}$}{$\SW B (0 , 1)$} algebra}
\label{sec:B(0,1)_IoM}
The algebra $\SW B (0 , 1)$ coincides with the $\svir$. 
The associated commuting family of local IoMs forms the quantum $\N = 1$ super-KdV hierarchy. 
This hierarchy has been studied in terms of $\svir$ modes \cite{yamanaka1988}, 
as the integrable structure underlying perturbations of $\N = 1$ superconformal minimal models \cite{mathieu1990}, 
and in terms of the component currents $T$ and $G$ \cite{alfimov2015,tanabe2026}. 
Since the affine root is even, we determine the densities from mutual commutativity \eqref{eq:conserved_law}. 
Up to total superderivatives, we choose the first three densities with the following normalizations: 
\begin{align}
  \widehat{J}_{\frac{3}{2}} &\coloneqq \widehat{G} , \\
  \widehat{J}_{\frac{7}{2}} &\coloneqq \; \normord{D \widehat{G} \, \widehat{G}} \, , \\
  \widehat{J}_{\frac{11}{2}} &\coloneqq \; \normord{D \widehat{G} \normord{D \widehat{G} \, \widehat{G}} \;} 
  + \; a_1 \normord{D^3 \widehat{G} \, D^2 \widehat{G}} . 
\end{align}
\par
Imposing mutual commutativity of $I_3$ and $I_5$ determines 
\begin{equation}
  a_1 = \frac{1}{12} \left( \! 6 \beta^2 + \frac{6}{\beta^2} - 19 \! \right) 
  = - \frac{c + 2}{6} . 
\end{equation}
This gives the first three local IoMs of the quantum $\N = 1$ super-KdV hierarchy in our superfield conventions. 

\subsection[$\SW B (1 , 1)$ algebra]{\texorpdfstring{$\bm{\SW B (1 , 1)}$}{$\SW B (1 , 1)$} algebra}
\label{sec:B(1,1)_IoM}
The first higher-spin conserved current of the $B (1 , 1)^{(1)}$ Toda theory was constructed in \cite{penati1992}. 
Here we determine $I_3$ using the commutativity condition \eqref{eq:integrable_perturbation}, 
since the affine root is odd. 
Up to total superderivatives, we choose the first three densities with the following normalizations: 
\begin{align}
  \widehat{J}_{\frac{3}{2}} &\coloneqq \widehat{G} , \\
  \widehat{J}_{\frac{7}{2}} &\coloneqq \; \normord{\widehat{W}_2 \widehat{G}} 
  + \; a_1 \normord{D \widehat{G} \, \widehat{G}} \, , \\
  \widehat{J}_{\frac{11}{2}} &\coloneqq \; \normord{\widehat{W}_2 \normord{\widehat{W}_2 \widehat{G}} \;} 
  + \; b_1 \normord{\widehat{W}_2 \normord{D \widehat{G} \, \widehat{G}} \;} 
  + \; b_2 \normord{D \widehat{G} \normord{D \widehat{G} \, \widehat{G}} \;} \notag \\
  &\quad + b_3 \normord{D^2 \widehat{W}_2 \, D \widehat{W}_2} 
  + \; b_4 \normord{D^2 \widehat{W}_2 \, D^2 \widehat{G}} 
  + \; b_5 \normord{D^3 \widehat{G} \, D^2 \widehat{G}} . 
\end{align}
\par
The commutativity condition $[Q_0 , I_3] = 0$ fixes 
\begin{equation}
  a_1 = \frac{\beta^2 - 5}{2 \left( \beta^2 - 2 \right)} . 
\end{equation}
With $I_3$ fixed, mutual commutativity, $[I_3 , I_5] = 0$, determines the remaining coefficients: 
\begin{align}
  b_1 &= - \frac{i \left( 2 \beta^4 - 14 \beta^2 + 5 \right)}{3 \beta \left( \beta^2 - 2 \right)} , \quad 
  b_2 = - \frac{4 \beta^8 - 56 \beta^6 + 198 \beta^4 - 143 \beta^2 + 28}
  {36 \beta^2 \left( \beta^2 - 2 \right)^2} , \quad 
  b_3 = - \frac{1}{2 \beta^2} - 1 , \notag \\
  b_4 &= \frac{i \left( \beta^6 - 5 \beta^4 - 24 \beta^2 + 10 \right)}{6 \beta^3 \left( \beta^2 - 2 \right)} , \quad 
  b_5 = - \frac{4 \beta^{10} - 42 \beta^8 + 247 \beta^6 - 643 \beta^4 + 437 \beta^2 - 84}
  {72 \beta^4 \left( \beta^2 - 2 \right)^2} . 
\end{align}
These densities give a commuting set of local IoMs $I_1$, $I_3$, and $I_5$ for $\SW B (1 , 1)$. 

\subsection[$\SW B (1 , 2)$ algebra]{\texorpdfstring{$\bm{\SW B (1 , 2)}$}{$\SW B (1 , 2)$} algebra}
\label{sec:B(1,2)_IoM}
The affine root of $B (1 , 2)^{(1)}$ is even. 
The corresponding screening charge therefore cannot be introduced 
within the purely odd screening framework used here. 
We therefore determine $\widehat{J}_{\frac{7}{2}}$ and 
$\widehat{J}_{\frac{11}{2}}$ simultaneously by imposing mutual commutativity of $I_3$ and $I_5$. 
Up to total superderivatives, we choose the first three densities with the following normalizations: 
\begin{align}
  \widehat{J}_{\frac{3}{2}} &\coloneqq \widehat{G} , \\
  \widehat{J}_{\frac{7}{2}} &\coloneqq \widehat{W}_{\frac{7}{2}} + a_1 \normord{\widehat{W}_2 \widehat{G}} 
  + \; a_2 \normord{D \widehat{G} \, \widehat{G}} \, , \\
  \widehat{J}_{\frac{11}{2}} &\coloneqq \; \normord{\widehat{W}_{\frac{7}{2}} \widehat{W}_2} 
  + \; b_1 \normord{\widehat{W}_{\frac{7}{2}} \, D \widehat{G}} 
  + \; b_2 \normord{\widehat{W}_2 \normord{\widehat{W}_2 \widehat{G}} \;} 
  + \; b_3 \normord{\widehat{W}_2 \normord{D \widehat{G} \, \widehat{G}} \;} \notag \\
  &\quad + b_4 \normord{D \widehat{G} \normord{D \widehat{G} \, \widehat{G}} \;} 
  + \; b_5 \normord{D^2 \widehat{W}_2 \, D \widehat{W}_2} 
  + \; b_6 \normord{D^2 \widehat{W}_2 \, D^2 \widehat{G}} 
  + \; b_7 \normord{D^3 \widehat{G} \, D^2 \widehat{G}} . 
\end{align}
\par
The equations enforcing mutual commutativity, $[I_3 , I_5] = 0$, admit several algebraic branches. 
For generic $\beta$, we obtain the following two rational solutions. 
The first rational solution is 
\begin{align}
  a_1 &= \frac{3 \left( \beta^2 - 6 \right) \left( 4 \beta^2 - 3 \right)}
  {\left( \beta^2 - 3 \right) \left( 2 \beta^2 - 1 \right)} , \quad 
  a_2 = - \frac{3 \left( \beta^2 - 6 \right) \left( 4 \beta^2 - 3 \right) \left( 2 \beta^4 + \beta^2 + 6 \right) 
  \left( 10 \beta^4 - 59 \beta^2 + 60 \right)}
  {4 \left( \beta^2 - 3 \right)^2 \left( 2 \beta^2 - 1 \right) \left( 4 \beta^4 - 29 \beta^2 + 24 \right)} , \\
  b_1 &= - \frac{10 \beta^2 + 15}{2 \beta^2 - 6} , \quad 
  b_2 = \frac{7 \beta^2 - 48}{2 \beta^2 - 6} , \quad 
  b_3 = - \frac{260 \beta^8 - 2752 \beta^6 + 3861 \beta^4 + 26460 \beta^2 - 28080}
  {4 \left( \beta^2 - 3 \right)^2 \left( 4 \beta^4 - 29 \beta^2 + 24 \right)} , \notag \\
  b_4 &= \frac{\left( 10 \beta^4 - 59 \beta^2 + 60 \right) 
  \left( 144 \beta^8 - 688 \beta^6 - 531 \beta^4 - 810 \beta^2 - 4104 \right)}
  {16 \left( \beta^2 - 3 \right)^3 \left( 4 \beta^4 - 29 \beta^2 + 24 \right)} , \notag \\
  b_5 &= - \frac{3 \left( 4 \beta^6 - 4 \beta^4 - 97 \beta^2 + 84 \right)}
  {8 \beta^2 \left( \beta^2 - 3 \right)} , \notag \\
  b_6 &= - \frac{360 \beta^{12} - 6300 \beta^{10} 
  + 38742 \beta^8 - 86061 \beta^6 - 3060 \beta^4 + 131760 \beta^2 - 77760}
  {8 \beta^2 \left( \beta^2 - 3 \right)^2 \left( 4 \beta^4 - 29 \beta^2 + 24 \right)} , \notag \\
  b_7 &= \frac{3 \left( 10 \beta^4 - 59 \beta^2 + 60 \right) 
  \left( 96 \beta^{12} - 1056 \beta^{10} + 3390 \beta^8 - 4181 \beta^6 + 8790 \beta^4 + 144 \beta^2 - 12096 \right)}
  {64 \beta^2 \left( \beta^2 - 3 \right)^3 \left( 4 \beta^4 - 29 \beta^2 + 24 \right)} . 
\end{align}
The second rational solution is 
\begin{align}
  a_1 &= - \frac{3 \left( \beta^2 + 4 \right)}{2 \left(\beta ^2-2\right)} , \quad 
  a_2 = -\frac{3 \left( \beta^2 + 4 \right) \left( 4 \beta^4 - 21 \beta^2 + 8 \right) 
  \left( 10 \beta^4 - 59 \beta^2 + 60 \right)}
  {16 \left( \beta^2 - 2 \right)^2 \left( 4 \beta^4 - 29 \beta^2 + 24 \right)} , \\[3mm]
  b_1 &= -\frac{5 \left( \beta^2 + 4 \right)}{4 \left( \beta^2 - 2 \right)} , \quad 
  b_2 = - \frac{22 \beta^4 - 17 \beta^2 + 36}{4 \left( \beta^2 - 2 \right) \left( 2 \beta^2 - 1 \right)} , \notag \\ 
  b_3 &= - \frac{200 \beta^{10} - 2766 \beta^8 + 15579 \beta^6 - 14464 \beta^4 + 18480 \beta^2 - 17280}
  {16 \left( \beta^2 - 2 \right)^2 \left( 2 \beta^2 - 1 \right) 
  \left( 4 \beta^4 - 29 \beta^2 + 24 \right)} , \notag \\
  b_4 &= \frac{\left( 10 \beta^4 - 59 \beta^2 + 60 \right) 
  \left( 144 \beta^{10} + 832 \beta^8 - 5769 \beta^6 - 1644 \beta^4 + 7360 \beta^2 - 6912 \right)}
  {128 \left( \beta^2 - 2 \right)^3 \left( 2 \beta^2 - 1 \right) 
  \left( 4 \beta^4 - 29 \beta^2 + 24 \right)} , \notag \\
  b_5 &= - \frac{48 \beta^8 - 274 \beta^6 - 93 \beta^4 + 616 \beta^2 - 336}
  {16 \beta^2 \left( \beta^2 - 2 \right) \left( 2 \beta^2 - 1 \right)} , \notag \\
  b_6 &= - \frac{720 \beta^{14} - 9620 \beta^{12} + 49356 \beta^{10} 
  - 109259 \beta^8 + 19524 \beta^6 + 151600 \beta^4 - 127680 \beta^2 + 23040}
  {32 \beta^2 \left( \beta^2 - 2 \right)^2 \left( 2 \beta^2 - 1 \right) 
  \left( 4 \beta^4 - 29 \beta^2 + 24 \right)} , \notag \\
  b_7 &= \frac{10 \beta^4 - 59 \beta^2 + 60}
  {512 \beta^2 \left( \beta^2 - 2 \right)^3 \left( 2 \beta^2 - 1 \right) 
  \left( 4 \beta^4 - 29 \beta^2 + 24 \right)} \notag \\
  &\quad \times \left( 288 \beta^{14} - 1008 \beta^{12} - 7426 \beta^{10} 
  + 28103 \beta^8 + 23304 \beta^6 - 75152 \beta^4 + 38656 \beta^2 - 21504 \right) . 
\end{align}
Both solutions satisfy $[I_3 , I_5] = 0$. 
Whether they extend to infinite commuting hierarchies remains an open question. 

\subsection[$\SW B (2 , 2)$ algebra]{\texorpdfstring{$\bm{\SW B (2 , 2)}$}{$\SW B (2 , 2)$} algebra}
\label{sec:B(2,2)_IoM}
Since the affine root of $B (2 , 2)^{(1)}$ is odd, 
we determine $I_3$ using the commutativity condition \eqref{eq:integrable_perturbation}. 
Up to total superderivatives, we choose the first three densities with the following normalizations: 
\begin{align}
  \widehat{J}_{\frac{3}{2}} &\coloneqq \widehat{G} , \\
  \widehat{J}_{\frac{7}{2}} &\coloneqq \widehat{W}_{\frac{7}{2}} + a_1 \normord{\widehat{W}_2 \widehat{G}} 
  + \; a_2 \normord{D \widehat{G} \, \widehat{G}} \, , \\
  \widehat{J}_{\frac{11}{2}} &\coloneqq \; \normord{\widehat{W}_4 \widehat{G}} 
  + \; b_1 \normord{\widehat{W}_{\frac{7}{2}} \widehat{W}_2} 
  + \; b_2 \normord{\widehat{W}_{\frac{7}{2}} \, D \widehat{G}} 
  + \; b_3 \normord{\widehat{W}_2 \normord{\widehat{W}_2 \widehat{G}} \;} 
  + \; b_4 \normord{\widehat{W}_2 \normord{D \widehat{G} \, \widehat{G}} \;} \notag \\
  &\quad + b_5 \normord{D \widehat{G} \normord{D \widehat{G} \, \widehat{G}} \;} 
  + \; b_6 \normord{D^2 \widehat{W}_2 \, D \widehat{W}_2} 
  + \; b_7 \normord{D^2 \widehat{W}_2 \, D^2 \widehat{G}} 
  + \; b_8 \normord{D^3 \widehat{G} \, D^2 \widehat{G}} . 
\end{align}
\par
We first impose the commutativity condition $\left[ Q_0 , I_3 \right] = 0$. 
For generic $\beta$, this condition fixes the coefficients in $\widehat{J}_{\frac{7}{2}}$ as 
\begin{equation}
  a_1 = \frac{9 i \beta \left( \beta^2 - 40 \right) \left( \beta^2 - 7 \right) \left( \beta^2 + 2 \right)}
  {2 \left( \beta^2 - 4 \right)} , \quad 
  a_2 = -\frac{3 \left( 4 \beta^2 - 25 \right) \left( \beta^2 - 7 \right) \left( \beta^2 + 2 \right) 
  \left( 2 \beta^4 + 3 \beta^2 - 8 \right)}{2 \left( \beta^2 - 4 \right)^2} . 
\end{equation}
Having fixed $I_3$ by the commutativity condition, we determine $I_5$ by imposing mutual commutativity,
$[I_3 , I_5] = 0$. 
This condition determines the remaining coefficients as 
\begin{align}
  b_1 &= \frac{\beta^2 \left( 17 \beta^2 + 60 \right) \left( 67 \beta^2 - 300 \right) 
  \left( 10 \beta^4 - 59 \beta^2 + 60 \right)}
  {\left( 4 \beta^2 - 25 \right) \left( \beta^2 - 40 \right) \left( 2 \beta^2 - 1 \right)} , \notag \\
  b_2 &= - \frac{i \beta  \left( 17 \beta^2 + 60 \right) \left( 67 \beta^2 - 300 \right) 
  \left( 130 \beta^8 - 1983 \beta^6 + 8831 \beta^4 - 12426 \beta^2 + 5400 \right)}
  {21 \left( \beta^2 - 4 \right) \left( 4 \beta^2 - 25 \right) 
  \left( \beta^2 - 40 \right) \left( 2 \beta^2 - 1 \right)} , \notag \\
  b_3 &= - \frac{3 i \beta^3 \left( 17 \beta^2 + 60 \right) \left( 67 \beta^2 - 300 \right) 
  \left( 10 \beta^8 - 207 \beta^6 + 1622 \beta^4 - 4701 \beta^2 + 2820 \right)}
  {2 \left( \beta^2 - 4 \right) \left( 4 \beta^2 - 25 \right) \left( 2 \beta^2 - 1 \right)} , \notag \\
  b_4 &= - \frac{\beta^2 \left( 67 \beta^2 - 300 \right)}
  {2 \left( \beta^2 - 4 \right)^2 \left( 4 \beta^2 - 25 \right) 
  \left( \beta^2 - 40 \right) \left( 2 \beta^2 - 1 \right)} 
  \big( 3670 \beta^{16} - 25355 \beta^{14} - 326910 \beta^{12} \notag \\
  &\quad + 2876213 \beta^{10} - 1694524 \beta^8 - 
  12888342 \beta^6 - 52823520 \beta^4 + 165924000 \beta^2 - 90720000 \big) , \notag \\
  b_5 &= \frac{i \beta  \left( 2 \beta^2 - 3 \right) \left( 17 \beta^2 + 60 \right)}
  {12 \left( \beta^2 - 4 \right)^3 \left( \beta^2 - 40 \right) \left( 2 \beta^2 - 1 \right)} 
  \big( 8110 \beta^{16} - 205769 \beta^{14} + 1992393 \beta^{12} \notag \\
  &\quad - 9350899 \beta^{10} + 23794133 \beta^8 - 44559720 \beta^6 
  + 87477456 \beta^4 - 105707520 \beta^2 + 43545600 \big) , \notag \\
  b_6 &= \frac{3 i \beta \left( \beta^2 - 5 \right) \left( 17 \beta^2 + 60 \right) 
  \left( 67 \beta^2 - 300 \right) \left( 30 \beta^8 - 373 \beta^6 + 1324 \beta^4 - 213 \beta^2 - 540 \right)}
  {2 \left( \beta^2 - 4 \right) \left( 4 \beta^2 - 25 \right) \left( 2 \beta^2 - 1 \right)} , \notag \\
  b_7 &= - \frac{\left( 67 \beta^2 - 300 \right)}{4 \left( \beta^2 - 4 \right)^2 \left( 4 \beta^2 - 25 \right) 
  \left( \beta^2 - 40 \right) \left( 2 \beta^2 - 1 \right)} \notag \\
  &\quad \times \big( 7860 \beta^{18} - 167572 \beta^{16} 
  + 1002253 \beta^{14} + 1894386 \beta^{12} - 34240575 \beta^{10} \notag \\
  &\quad + 56740924 \beta^8 + 144083000 \beta^6 
  - 135259200 \beta^4 - 343728000 \beta^2 + 259200000 \big) , \notag \\
  b_8 &= \frac{i \left( 2 \beta^2 - 3 \right) \left( 17 \beta^2 + 60 \right)}
  {24 \beta \left( \beta^2 - 4 \right)^3 \left( \beta^2 - 40 \right) \left( 2 \beta^2 - 1 \right)} \notag \\
  &\quad \times \big( 14070 \beta^{18} - 431923 \beta^{16} 
  + 5306526 \beta^{14} - 33227099 \beta^{12} + 110456676 \beta^{10} \notag \\
  &\quad - 187703442 \beta^8 + 192341904 \beta^6 - 325596960 \beta^4 + 444787200 \beta^2 - 186624000 \big) . 
\end{align}
This completes the construction of the local IoMs $I_1$, $I_3$, and $I_5$ for the four algebras considered here, 
with two rational solutions for $\SW B (1 , 2)$. 
\section{Conclusions and outlook}
\label{sec:conclusion}
In this paper, we extended the recursive construction of free-field realizations 
to $\N = 1$ SUSY $\W$-algebras associated with Lie superalgebras admitting purely odd simple root systems. 
At each step, one fermionic free superfield is added to the realization of the preceding algebra, 
and the zero commutation relation associated with the additional simple root is imposed. 
Applying this procedure to the $B (n - 1 , n)$ and $B (n , n)$ families, 
we obtained explicit free-field realizations of the first four algebras in the sequence 
\begin{equation}
  \begin{aligned}
    \begin{tikzpicture}[
        x=3.2cm, y=1.2cm,
        every node/.style={inner sep=3pt},
        >={Stealth}
      ]
      \node[anchor=east] at (-0.65,0)
        {$\SW B (n - 1 , n):$};
      \node[anchor=east] at (-0.65,-1)
        {$\SW B (n , n):$};
      \node (B01) at (0,0) {$\SW B (0 , 1)$};
      \node (B11) at (0.5,-1) {$\SW B (1 , 1)$};
      \node (B12) at (1,0) {$\SW B (1 , 2)$};
      \node (B22) at (1.5,-1) {$\SW B (2 , 2)$};
      \draw[->] (B01) -- (B11);
      \draw[->] (B11) -- (B12);
      \draw[->] (B12) -- (B22);
    \end{tikzpicture}
  \end{aligned}
  . 
\end{equation}
To the best of our knowledge, this is the first formulation of such a recursive construction for $\SW$-algebras. 
\par
We expressed the superconformal currents and central charges in terms of the Weyl and co-Weyl vectors 
and explicitly determined the additional generating $\W$-currents up to rank four. 
To the best of our knowledge, explicit free-field expressions for the generating currents of the $\SW B$-algebras 
were previously available only up to $\SW B (1 , 1)$. 
Our results provide new explicit realizations of $\SW B (1 , 2)$ and $\SW B (2 , 2)$. 
We also computed the conformal weights and $\W$-charges of free-field highest-weight states and expressed them 
in terms of Casimir invariants of the bosonic subalgebras evaluated at shifted momenta. 
Their invariance under the Weyl groups of the bosonic subalgebras 
provides a consistency check of the recursive construction. 
\par
We also constructed the local IoMs $I_1$, $I_3$, and $I_5$ for these algebras. 
Beyond the previously studied $I_3$ of the $B (1 , 1)^{(1)}$ Toda theory \cite{penati1992}, 
we obtained new results for $I_5$ of $\SW B (1 , 1)$ and for $I_3$ and $I_5$ of $\SW B (1 , 2)$ and $\SW B (2 , 2)$. 
For $\SW B (1 , 2)$, we found two rational branches, each giving a commuting pair $(I_3 , I_5)$. 
Whether these branches extend to infinite commuting hierarchies remains an open question. 
\par
For the $B (n , n)$ family, the affine root is odd, 
allowing an additional screening within the same fermionic-superfield formalism. 
In the cases studied here, we used this screening to fix $I_3$ and then determined $I_5$ from $[I_3 , I_5] = 0$. 
For the $B (n - 1 , n)$ family, by contrast, 
the affine root is even and falls outside the purely odd screening framework used here. 
We therefore determined the local densities directly from mutual commutativity. 
\par
Several directions remain open. 
It would be desirable to extend the recursive construction 
to arbitrary rank and clarify its relation to quantum DS reduction through the screening kernels. 
We have also obtained free-field realizations for the $D (n , n)$ and $D (n + 1 , n)$ families, 
which admit purely odd simple root systems; their detailed analysis is in progress. 
Other possible extensions include $\N \geq 2$ SUSY $\W$-algebras 
and non-SUSY $\W$-algebras associated with Lie superalgebras 
whose simple root systems contain both even and odd roots. 
It would also be interesting to examine the special values of $\beta$ at which our expressions become singular, 
distinguishing normalization singularities from additional null relations or degenerations of the generating set. 
\par
Finally, the explicit $\W$-charges and local IoMs obtained here 
provide a basis for investigating the SUSY ODE/IM correspondence 
for the $\SW B (n - 1 , n)$ and $\SW B (n , n)$ families. 
A natural next step is to compute the local IoM eigenvalues, identify the associated linear problems, 
and compare their WKB periods with these eigenvalues. 
The role of Langlands-dual algebras in the ordinary ODE/IM correspondence 
also motivates exploring analogous structures for Lie superalgebras. 
Superopers and their connections to Gaudin models, Bethe ansatz equations, and $q q$-systems 
have been studied in \cite{zeitlin2015,zeitlin2025}. 
Clarifying their relation to the ODE/IM correspondence 
for CFTs with the $\SW$-symmetries considered here may shed light on Langlands-type dualities for Lie superalgebras. 

\acknowledgments
The author is deeply grateful to K. Ito for valuable discussions, a careful reading of the manuscript, 
and his notes on Lie superalgebras, 
and to S. Miyazaki for numerous discussions on recursive constructions and integrable perturbations. 
The author thanks S. Komata for visiting our institute and for valuable discussions, 
J. Ekstrand for providing the Mathematica package \texttt{Lambda} and helpful discussions, 
and D. Ide and W. Kono for helpful discussions. 
The author also thanks T. Yano and S. Kudo for discussions 
during the 9th Seminar of Physics and Mathematics Advanced. 
\par
The author used ChatGPT (OpenAI) for assistance with coding and language editing 
and takes full responsibility for the content of this work. 

\appendix
\section{\boldmath Higher-spin currents of $\SW B (2 , 2)$}
\label{sec:SWB22}
\begingroup
\setlength{\jot}{2pt}
In this appendix, we collect the explicit expressions for the higher-spin generating currents 
$\widehat{W}_{\frac{7}{2}}$ and $\widehat{W}_4$ in the free-field realization 
of the $\SW B (2 , 2)$ algebra constructed in section~\ref{sec:SWB(2,2)}. 
\par
For the spin-$\frac{7}{2}$ current, we take the ansatz 
\begin{small}
\begin{align}
  \widehat{W}_{\frac{7}{2}} &= \; c_1 \widehat{W}_{\frac{7}{2}}^{B (1 ,\, 2)} 
  + c_2 \normord{\widehat{W}_2^{B (1 ,\, 2)} \, \widehat{G}^{B (1 ,\, 2)}} 
  + \; c_3 \normord{D \widehat{G}^{B (1 ,\, 2)} \, \widehat{G}^{B (1 ,\, 2)}} 
  + \; c_4 \, D^3 \widehat{W}_2^{B (1 ,\, 2)} 
  + c_5 \, D^4 \widehat{G}^{B (1 ,\, 2)} \notag \\
  &\quad + c_6 \normord{\; \normord{\widehat{W}_2^{B (1 ,\, 2)} \, 
  D \widehat{\varPsi}_1} \widehat{\varPsi}_1} 
  + \; c_7 \normord{D^2 \widehat{W}_2^{B (1 ,\, 2)} \, \widehat{\varPsi}_1} 
  + \; c_8 \normord{D \widehat{W}_2^{B (1 ,\, 2)} \, D \widehat{\varPsi}_1} 
  + \; c_9 \normord{\widehat{W}_2^{B (1 ,\, 2)} \, D^2 \widehat{\varPsi}_1} \notag \\
  &\quad + c_{10} \normord{\; \normord{D \widehat{G}^{B (1 ,\, 2)} \, D \widehat{\varPsi}_1} \widehat{\varPsi}_1} 
  + \; c_{11} \normord{\; \normord{\widehat{G}^{B (1 ,\, 2)} \, D^2 \widehat{\varPsi}_1} \widehat{\varPsi}_1} 
  + \; c_{12} \normord{\; \normord{\widehat{G}^{B (1 ,\, 2)} \, D \widehat{\varPsi}_1} D \widehat{\varPsi}_1} 
  \notag \\
  &\quad + c_{13} \normord{D^3 \widehat{G}^{B (1 ,\, 2)} \, \widehat{\varPsi}_1} 
  + \; c_{14} \normord{D^2 \widehat{G}^{B (1 ,\, 2)} \, D \widehat{\varPsi}_1} 
  + \; c_{15} \normord{D \widehat{G}^{B (1 ,\, 2)} \, D^2 \widehat{\varPsi}_1} 
  + \; c_{16} \normord{\widehat{G}^{B (1 ,\, 2)} \, D^3 \widehat{\varPsi}_1} \notag \\
  &\quad + c_{17} \normord{\; \normord{\; \normord{D \widehat{\varPsi}_1 \, 
  D \widehat{\varPsi}_1} D \widehat{\varPsi}_1} \widehat{\varPsi}_1} 
  + \; c_{18} \normord{\; \normord{D^3 \widehat{\varPsi}_1 \, D \widehat{\varPsi}_1} \widehat{\varPsi}_1} 
  + \; c_{19} \normord{\; \normord{D^2 \widehat{\varPsi}_1 \, D \widehat{\varPsi}_1} D \widehat{\varPsi}_1} \notag \\
  &\quad + c_{20} \normord{D^5 \widehat{\varPsi}_1 \, \widehat{\varPsi}_1} 
  + \; c_{21} \normord{D^4 \widehat{\varPsi}_1 \, D \widehat{\varPsi}_1} 
  + \; c_{22} \normord{D^3 \widehat{\varPsi}_1 \, D^2 \widehat{\varPsi}_1} 
  + \; c_{23} \, D^6 \widehat{\varPsi}_1 . 
\end{align}
\end{small}
We choose the overall normalization so that $c_1$ takes the value displayed below. 
We obtain 
\begin{small}
\begin{align}
  c_1 &= \frac{\left( 4 \beta^2 - 25 \right) \left( \beta^2 - 40 \right) \left( 2 \beta^2 - 1 \right)}
  {10 \beta^4 - 59 \beta^2 + 60} , \quad 
  c_2 = - \frac{3 \left( \beta^2 - 40 \right) \left( 10 \beta^6 - 101 \beta^4 + 179 \beta^2 + 160 \right)}
  {2 \left( 10 \beta^4 - 59 \beta^2 + 60 \right)} , \notag \\
  c_3 &= - \frac{\left( \beta^2 - 40 \right) 
  \left( 26 \beta^8 - 311 \beta^6 + 1054 \beta^4 - 1384 \beta^2 + 1200 \right)}
  {2 \beta^2 \left( 10 \beta^4 - 59 \beta^2 + 60 \right)} , \notag \\
  c_4 &= - \frac{\left( 3 \beta^2 - 5 \right) 
  \left( 104 \beta^{10} - 1174 \beta^8 + 3621 \beta^6 + 3196 \beta^4 - 36416 \beta^2 + 15360 \right)}
  {8 \beta^2 \left( 4 \beta^4 - 29 \beta^2 + 24 \right)} , \notag \\
  c_5 &= \frac{3 \left( 3 \beta^2 - 5 \right) 
  \left( 12 \beta^8 - 151 \beta^6 + 520 \beta^4 + 44 \beta^2 - 2048 \right)}{8 \beta^4 - 58 \beta^2 + 48} , \quad 
  c_6 = - \frac{3}{2} \left( \beta^2 - 9 \right) \left( \beta^2 - 40 \right) , \notag \\
  c_7 &= c_8 = - \frac{i \left( 4 \beta^2 - 25 \right) \left( \beta^2 - 40 \right)}{2 \beta} , \quad 
  c_9 = \frac{i \left( \beta^2 - 40 \right) \left( 3 \beta^4 - 23 \beta^2 + 8 \right)}{2 \beta} , \notag \\ 
  c_{10} &= \frac{3}{2} \left( 9 \beta^6 - 63 \beta^4 + 74 \beta^2 + 10 \right) , \quad 
  c_{11} = \frac{9}{2} \left( \beta^2 - 7 \right) \left( \beta^2 + 2 \right) \left( 3 \beta^2 - 5 \right) , \notag \\
  c_{12} &= \frac{9}{4} \left( 6 \beta^6 - 43 \beta^4 + 91 \beta^2 - 60 \right) , \quad 
  c_{13} = c_{14} = \frac{i \left( \beta^2 - 4 \right) 
  \left( \beta^2 - 40 \right) \left( 3 \beta^2 - 5 \right)}{4 \beta} , \notag \\
  c_{15} &= - \frac{i \left( \beta^2 - 4 \right) \left( 3 \beta^2 - 5 \right) 
  \left( 9 \beta^4 - 46 \beta^2 - 86 \right)}{2 \beta} , \quad 
  c_{16} = - \frac{3 i \left( \beta^2 - 4 \right) \left( 3 \beta^2 - 5 \right) 
  \left( 6 \beta^4 - 31 \beta^2 - 44 \right)}{4 \beta} , \notag \\
  c_{17} &= \frac{9}{4} \left( 6 \beta^6 - 43 \beta^4 + 91 \beta^2 - 60 \right) , \quad 
  c_{18} = c_{19} = - \frac{9 i \left( \beta^2 - 4 \right)^2 
  \left( 6 \beta^4 - 19 \beta^2 + 15 \right)}{4 \beta} , \notag \\
  c_{20} &= - \frac{3 \left( \beta^2 - 4 \right) \left( \beta^2 + 5 \right) 
  \left( 2 \beta^2 - 3 \right) \left( 3 \beta^2 - 5 \right)}{2 \beta^2} , \quad 
  c_{21} = \frac{9 \left( \beta^2 - 4 \right) \left( \beta^2 - 10 \right) 
  \left( 2 \beta^2 - 3 \right) \left( 3 \beta^2 - 5 \right)}{4 \beta^2} , \notag \\
  c_{22} &= - \frac{9 \left( \beta^2 - 4 \right) \left( 2 \beta^2 - 3 \right) 
  \left( 3 \beta^2 - 5 \right) \left( \beta^4 - 6 \beta^2 - 4 \right)}{4 \beta^2} , \quad 
  c_{23} = \frac{i \left( \beta^2 - 4 \right)^2 \left( 2 \beta^2 - 3 \right) 
  \left( 3 \beta^2 - 5 \right) \left( 7 \beta^2 + 20 \right)}{8 \beta^3} . 
\end{align}
\end{small}
\par
For the spin-$4$ current, we take the ansatz
\begin{small}
\begin{align}
  \widehat{W}_4 &= \; d_1 \normord{\widehat{W}_2^{B (1 ,\, 2)} \widehat{W}_2^{B (1 ,\, 2)}} 
  + \; d_2 \normord{D \widehat{W}_2^{B (1 ,\, 2)} \, \widehat{G}^{B (1 ,\, 2)}} 
  + \; d_3 \normord{\widehat{W}_2^{B (1 ,\, 2)} \, D \widehat{G}^{B (1 ,\, 2)}} \notag \\
  &\quad + d_4 \normord{D^2 \widehat{G}^{B (1 ,\, 2)} \, \widehat{G}^{B (1 ,\, 2)}} 
  + \; d_5 \normord{D \widehat{G}^{B (1 ,\, 2)} \, D \widehat{G}^{B (1 ,\, 2)}} 
  + \; d_6 \, D \widehat{W}_{\frac{7}{2}}^{B (1 ,\, 2)} 
  + d_7 \, D^4 \widehat{W}_2^{B (1 ,\, 2)} \notag \\
  &\quad + d_8 \, D^5 \widehat{G}^{B (1 ,\, 2)} 
  + d_9 \normord{\widehat{W}_{\frac{7}{2}}^{B (1 ,\, 2)} \widehat{\varPsi}_1} 
  + \; d_{10} \normord{\; \normord{\widehat{W}_2^{B (1 ,\, 2)} \widehat{G}^{B (1 ,\, 2)}} \widehat{\varPsi}_1} 
  \notag \\
  &\quad + d_{11} \normord{\; \normord{D \widehat{W}_2^{B (1 ,\, 2)} \, D \widehat{\varPsi}_1} \widehat{\varPsi}_1} 
  + \; d_{12} \normord{\; \normord{\widehat{W}_2^{B (1 ,\, 2)} \, D^2 \widehat{\varPsi}_1} \widehat{\varPsi}_1} 
  + \; d_{13} \normord{\; \normord{\widehat{W}_2^{B (1 ,\, 2)} \, D \widehat{\varPsi}_1} D \widehat{\varPsi}_1} \notag \\
  &\quad + d_{14} \normord{D^3 \widehat{W}_2^{B (1 ,\, 2)} \, \widehat{\varPsi}_1} 
  + \; d_{15} \normord{D^2 \widehat{W}_2^{B (1 ,\, 2)} \, D \widehat{\varPsi}_1} 
  + \; d_{16} \normord{D \widehat{W}_2^{B (1 ,\, 2)} \, D^2 \widehat{\varPsi}_1} 
  + \; d_{17} \normord{\widehat{W}_2^{B (1 ,\, 2)} \, D^3 \widehat{\varPsi}_1} \notag \\
  &\quad + d_{18} \normord{\; \normord{D \widehat{G}^{B (1 ,\, 2)} \, \widehat{G}^{B (1 ,\, 2)}} \widehat{\varPsi}_1} 
  + \; d_{19} \normord{\; \normord{\; \normord{\widehat{G}^{B (1 ,\, 2)} \, 
  D \widehat{\varPsi}_1} D \widehat{\varPsi}_1} \widehat{\varPsi}_1} 
  + \; d_{20} \normord{\; \normord{D^2 \widehat{G}^{B (1 ,\, 2)} \, D \widehat{\varPsi}_1} \widehat{\varPsi}_1} 
  \notag \\
  &\quad 
  + d_{21} \normord{\; \normord{D \widehat{G}^{B (1 ,\, 2)} \, D^2 \widehat{\varPsi}_1} \widehat{\varPsi}_1} 
  + \; d_{22} \normord{\; \normord{D \widehat{G}^{B (1 ,\, 2)} \, D \widehat{\varPsi}_1} D \widehat{\varPsi}_1} 
  + \; d_{23} \normord{\; \normord{\widehat{G}^{B (1 ,\, 2)} \, D^3 \widehat{\varPsi}_1} \widehat{\varPsi}_1} \notag \\
  &\quad + d_{24} \normord{\; \normord{\widehat{G}^{B (1 ,\, 2)} \, 
  D^2 \widehat{\varPsi}_1} D \widehat{\varPsi}_1} 
  + \; d_{25} \normord{D^4 \widehat{G}^{B (1 ,\, 2)} \, \widehat{\varPsi}_1} 
  + \; d_{26} \normord{D^3 \widehat{G}^{B (1 ,\, 2)} \, D \widehat{\varPsi}_1} \notag \\
  &\quad + d_{27} \normord{D^2 \widehat{G}^{B (1 ,\, 2)} \, D^2 \widehat{\varPsi}_1} 
  + \; d_{28} \normord{D \widehat{G}^{B (1 ,\, 2)} \, D^3 \widehat{\varPsi}_1} 
  + \; d_{29} \normord{\widehat{G}^{B (1 ,\, 2)} \, D^4 \widehat{\varPsi}_1} \notag \\
  &\quad + d_{30} \normord{\; \normord{\; \normord{D^2 \widehat{\varPsi}_1 \, D \widehat{\varPsi}_1} D \widehat{\varPsi}_1} \widehat{\varPsi}_1} 
  + \; d_{31} \normord{\; \normord{\; \normord{D \widehat{\varPsi}_1 \, D \widehat{\varPsi}_1} D \widehat{\varPsi}_1} D \widehat{\varPsi}_1} 
  + \; d_{32} \normord{\; \normord{D^4 \widehat{\varPsi}_1 \, D \widehat{\varPsi}_1} \widehat{\varPsi}_1} \notag \\
  &\quad + d_{33} \normord{\; \normord{D^3 \widehat{\varPsi}_1 \, D^2 \widehat{\varPsi}_1} \widehat{\varPsi}_1} 
  + \; d_{34} \normord{\; \normord{D^3 \widehat{\varPsi}_1 \, D \widehat{\varPsi}_1} D \widehat{\varPsi}_1} 
  + \; d_{35} \normord{D^6 \widehat{\varPsi}_1 \, \widehat{\varPsi}_1} \notag \\
  &\quad 
  + d_{36} \normord{D^5 \widehat{\varPsi}_1 \, D \widehat{\varPsi}_1} 
  + \; d_{37} \normord{D^4 \widehat{\varPsi}_1 \, D^2 \widehat{\varPsi}_1} 
  + \; d_{38} \normord{D^3 \widehat{\varPsi}_1 \, D^3 \widehat{\varPsi}_1} 
  + \; d_{39} \, D^7 \widehat{\varPsi}_1 . 
\end{align}
\end{small}
We fix the composite-field ambiguity by setting $d_{10} = 0$ 
and choose the overall normalization so that $d_9$ takes the value displayed below. 
We obtain 
\begin{small}
\begin{align}
  d_1 &= - \frac{i}{6} \beta \left( 5 \beta^2 - 18 \right) 
  \left( 17 \beta^2 + 60 \right) \left( 67 \beta^2 - 300 \right) , \notag \\
  d_2 &= - \frac{2 i \beta \left( 67 \beta^2 - 300 \right) 
  \left( 450 \beta^{10} - 3830 \beta^8 + 9187 \beta^6 + 13374 \beta^4 - 91620 \beta^2 + 97200 \right)}
  {7 \left( 10 \beta^4 - 59 \beta^2 + 60 \right)} , \notag \\
  d_3 &= \frac{2 i \beta \left( 67 \beta^2 - 300 \right) 
  \left( 1800 \beta^{10} - 20675 \beta^8 + 49621 \beta^6 + 120654 \beta^4 - 525240 \beta^2 + 388800 \right)}
  {21 \left( 10 \beta^4 - 59 \beta^2 + 60 \right)} , \notag \\
  d_4 &= \frac{i \beta}{14 \left( 4 \beta^4 - 29 \beta^2 + 24 \right)} 
  \big( 36000 \beta^{14} - 881484 \beta^{12} + 7180901 \beta^{10} - 18648787 \beta^8 \notag \\
  &\qquad\qquad\qquad - 28608648 \beta^6 + 175118688 \beta^4 - 152349120 \beta^2 + 34214400 \big) , 
  \notag \\
  d_5 &= - \frac{i \beta}{84 \left( 4 \beta^4 - 29 \beta^2 + 24 \right)} 
  \big( 373680 \beta^{14} - 7329240 \beta^{12} + 52387055 \beta^{10} - 136475671 \beta^8 \notag \\
  &\qquad\qquad\qquad - 123742428 \beta^6 + 1243522080 \beta^4 - 1909232640 \beta^2 + 796262400 \big) , 
  \notag \\
  d_6 &= \frac{2 i \beta \left( 17 \beta^2 + 60 \right) \left( 67 \beta^2 - 300 \right) 
  \left( 10 \beta^6 - 70 \beta^4 + 157 \beta^2 - 120 \right)}
  {7 \left( 10 \beta^4 - 59 \beta^2 + 60 \right)} , \notag \\
  d_7 &= - \frac{i \left( 67 \beta^2 - 300 \right) 
  \left( 4310 \beta^{12} - 38245 \beta^{10} + 129584 \beta^8 
  - 62284 \beta^6 - 847824 \beta^4 + 1206720 \beta^2 - 172800 \right)}
  {28 \beta \left( 10 \beta^4 - 59 \beta^2 + 60 \right)} , \notag \\
  d_8 &= \frac{i}{56 \beta \left( 4 \beta^4 - 29 \beta^2 + 24 \right)} \big( 828960 \beta^{16} 
  - 14708356 \beta^{14} + 90023627 \beta^{12} - 172353457 \beta^{10} \notag \\
  &\qquad\qquad\qquad - 346565236 \beta^8+ 1804809408 \beta^6 
  - 2247471360 \beta^4 + 889228800 \beta^2 - 41472000 \big) , \notag \\
  d_9 &= \beta^2 \left( 17 \beta^2 + 60 \right) \left( 67 \beta^2 - 300 \right) , \quad 
  d_{10} = 0 , \quad 
  d_{11} = - \frac{i}{14} \beta \left( 4 \beta^2 - 3 \right) \left( 67 \beta^2 - 300 \right) 
  \left( 45 \beta^4 - 88 \beta^2 - 480 \right) , \notag \\
  d_{12} &= \frac{6 i}{7} \beta \left( 4 \beta^2 - 3 \right) \left( 67 \beta^2 - 300 \right) 
  \left( 5 \beta^4 - 23 \beta^2 - 100 \right) , \quad 
  d_{13} = \frac{i}{7} \beta \left( 4 \beta^2 - 3 \right) \left( 67 \beta^2 - 300 \right) 
  \left( 30 \beta^4 - 19 \beta^2 - 180 \right) , \notag \\
  d_{14} &= - \frac{1}{14} \left( \beta^2 - 4 \right) \left( 4 \beta^2 - 3 \right) 
  \left( 17 \beta^2 + 60 \right) \left( 67 \beta^2 - 300 \right) , \quad 
  d_{15} = \frac{5}{14} \left( \beta^2 - 4 \right) \left( 4 \beta^2 - 3 \right) 
  \left( 17 \beta^2 + 60 \right) \left( 67 \beta^2 - 300 \right) , \notag \\
  d_{16} &= - \frac{1}{14} \left( \beta^2 - 4 \right) \left( 4 \beta^2 - 3 \right) 
  \left( 67 \beta^2 - 300 \right) \left( 45 \beta^4 - 173 \beta^2 - 780 \right) , \notag \\
  d_{17} &= \frac{2}{7} \left( \beta^2 - 4 \right) \left( 4 \beta^2 - 3 \right) 
  \left( 67 \beta^2 - 300 \right) \left( 15 \beta^4 - 52 \beta^2 - 240 \right) , \notag \\
  d_{18} &= - \frac{3 \beta^2 \left( 4 \beta^2 - 3 \right) \left( 67 \beta^2 - 300 \right) 
  \left( 30 \beta^8 - 296 \beta^6 + 493 \beta^4 + 2064 \beta^2 - 5040 \right)}
  {4 \beta^4 - 29 \beta^2 + 24} , \notag \\
  d_{19} &= - \frac{9}{2} \beta^2 \left( 4 \beta^2 - 3 \right) \left( 67 \beta^2 - 300 \right) 
  \left( 5 \beta^4 - 6 \beta^2 - 40 \right) , \notag \\
  d_{20} &= \frac{i}{7} \beta \left( 4 \beta^2 - 3 \right) 
  \left( 1125 \beta^8 - 3597 \beta^6 + 7808 \beta^4 - 67320 \beta^2 + 122400 \right) , \notag \\
  d_{21} &= \frac{3 i}{7} \beta \left( \beta^2 - 4 \right) \left( 4 \beta^2 - 3 \right) 
  \left( 2220 \beta^6 - 13861 \beta^4 - 11980 \beta^2 + 107400 \right) , \notag \\
  d_{22} &= - \frac{5 i}{14} \beta \left( 4 \beta^2 - 3 \right) 
  \left( 1557 \beta^8 - 14364 \beta^6 + 27640 \beta^4 + 79728 \beta^2 - 233280 \right) , \notag \\
  d_{23} &= \frac{3 i}{14} \beta \left( 4 \beta^2 - 3 \right) 
  \left( 6285 \beta^8 - 66354 \beta^6 + 131936 \beta^4 + 429600 \beta^2 - 1209600 \right) , \notag \\
  d_{24} &= - \frac{3 i}{7} \beta \left( 4 \beta^2 - 3 \right) 
  \left( 375 \beta^8 - 1869 \beta^6 - 1544 \beta^4 + 36360 \beta^2 - 79200 \right) , \notag \\
  d_{25} &= \frac{\left( 4 \beta^2 - 3 \right) \left( 67 \beta^2 - 300 \right) 
  \left( 3390 \beta^{10} - 31087 \beta^8 + 37158 \beta^6 + 228688 \beta^4 - 437760 \beta^2 + 28800 \right)}
  {28 \left( 4 \beta^4 - 29 \beta^2 + 24 \right)} , \notag \\
  d_{26} &= - \frac{1}{14} \left( 4 \beta^2 - 3 \right) \left( 67 \beta^2 - 300 \right) 
  \left( 60 \beta^6 - 13 \beta^4 - 1040 \beta^2 + 1200 \right) , \notag \\
  d_{27} &= \frac{1}{7} \left( 4 \beta^2 - 3 \right) 
  \left( 1125 \beta^{10} - 16137 \beta^8 + 59938 \beta^6 + 33008 \beta^4 - 393120 \beta^2 + 230400 \right) , 
  \notag \\
  d_{28} &= - \frac{1}{14} \left( 4 \beta^2 - 3 \right) 
  \left( 7785 \beta^{10} - 111000 \beta^8 + 463222 \beta^6 - 22600 \beta^4 - 3545760 \beta^2 + 5385600 \right) , 
  \notag \\
  d_{29} &= - \frac{3}{14} \left( 4 \beta^2 - 3 \right) 
  \left( 750 \beta^{10} - 9753 \beta^8 + 32935 \beta^6 + 37772 \beta^4 - 308880 \beta^2 + 273600 \right) , \notag \\
  d_{30} &= \frac{15 i}{14} \beta \left( \beta^2 - 4 \right) \left( 2 \beta^2 - 3 \right) 
  \left( 4 \beta^2 - 3 \right) \left( 519 \beta^4 - 694 \beta^2 - 3960 \right) , \notag \\
  d_{31} &= - \frac{15 i}{56} \beta \left( \beta^2 - 4 \right) \left( 2 \beta^2 - 3 \right) 
  \left( 4 \beta^2 - 3 \right) \left( 519 \beta^4 - 694 \beta^2 - 3960 \right) , \notag \\
  d_{32} &= \frac{9}{14} \left( \beta^2 - 4 \right) \left( 2 \beta^2 - 3 \right) 
  \left( 4 \beta^2 - 3 \right) \left( 125 \beta^6 - 268 \beta^4 + 120 \beta^2 - 2400 \right) , \notag \\
  d_{33} &= \frac{3}{14} \left( \beta^2 - 4 \right) \left( 2 \beta^2 - 3 \right) 
  \left( 4 \beta^2 - 3 \right) \left( 1845 \beta^6 - 12242 \beta^4 - 6640 \beta^2 + 93600 \right) , \notag \\
  d_{34} &= - \frac{15}{28} \left(\beta^2 - 4 \right)^2 \left( 2 \beta^2 - 3 \right) 
  \left( 4 \beta^2 - 3 \right) \left( 519 \beta^4 - 694 \beta^2 - 3960 \right) , \notag \\
  d_{35} &= - \frac{5 i \left( \beta^2 - 4 \right) \left( 2 \beta^2 - 3 \right) \left( 4 \beta^2 - 3 \right) 
  \left( 1701 \beta^6 - 1972 \beta^4 - 16200 \beta^2 + 7200 \right)}{28 \beta} , \notag \\
  d_{36} &= \frac{15 i \left( \beta^2 - 4 \right) \left( 2 \beta^2 - 3 \right) \left( 4 \beta^2 - 3 \right) 
  \left( 1110 \beta^6 - 1333 \beta^4 - 10080 \beta^2 + 3600 \right)}{14 \beta} , \notag \\
  d_{37} &= - \frac{9 i \left( \beta^2 - 4 \right) \left( 2 \beta^2 - 3 \right) \left( 4 \beta^2 - 3 \right) 
  \left( 125 \beta^8 - 1128 \beta^6 + 917 \beta^4 + 7920 \beta^2 - 8400 \right)}{14 \beta} , \notag \\
  d_{38} &= \frac{15 i \left( \beta^2 - 4 \right) \left( 2 \beta^2 - 3 \right) \left( 4 \beta^2 - 3 \right) 
  \left( 519 \beta^8 - 1645 \beta^6 + 4708 \beta^4 + 5816 \beta^2 - 84960 \right)}{56 \beta} , \notag \\
  d_{39} &= \frac{3 \left( \beta^2 - 4 \right) \left( 2 \beta^2 - 3 \right) \left( 4 \beta^2 - 3 \right) 
  \left( 4175 \beta^8 - 28417 \beta^6 - 19700 \beta^4 + 243600 \beta^2 - 24000 \right)}{56 \beta^2} . 
\end{align}
\end{small}
\endgroup

\bibliographystyle{JHEP}
\bibliography{SWB}
\end{document}